\documentclass[preprint,review,11pt]{elsarticle}

\usepackage{amssymb}
\usepackage{lineno}
\usepackage[table,xcdraw]{xcolor}
\usepackage{ccaption}
\usepackage{float}
\usepackage[numbers]{natbib}
\usepackage{amsthm}
\usepackage{mathrsfs,amsmath}
\usepackage{amsbsy}
\usepackage{graphicx}
\usepackage{subcaption}
\usepackage{siunitx}
\usepackage[english]{babel}
\usepackage{supertabular}
\usepackage{multirow}
\usepackage{bm} 
\usepackage[nameinlink,noabbrev]{cleveref}
\usepackage{array}
\usepackage{longtable}
\usepackage{booktabs}
\usepackage[normalem]{ulem}
\usepackage{tabularray}
\usepackage{xfrac,bigints}

\graphicspath{{images/}}

\newcommand{\rev}[1]{{\color{black}#1}}
\newcommand{\revv}[1]{{\color{black}#1}}

\begin{document}

\journal{}
\makeatletter
\def\ps@pprintTitle{%
  \let\@oddhead\@empty
  \let\@evenhead\@empty
  \let\@oddfoot\@empty
  \let\@evenfoot\@oddfoot
}
\makeatother

\begin{frontmatter}

\title{Flow control of three-dimensional cylinders transitioning to turbulence via multi-agent reinforcement learning}

\author[KTH]{P. Suárez}
\ead{polsm@kth.se}
\author[KTH]{F. Alcántara-Ávila }
\author[INDEPE]{J. Rabault}
\author[BSC]{A. Miró}
\author[DELFT]{B. Font}
\author[BSC]{O. Lehmkuhl}
\author[KTH]{R. Vinuesa}
\ead{rvinuesa@mech.kth.se}

\address[KTH]{FLOW, Engineering Mechanics, KTH Royal Institute of Technology, Stockholm, Sweden}
\address[INDEPE]{Independent researcher, Oslo, Norway}
\address[BSC]{Barcelona Supercomputing Center, Barcelona, Spain}
\address[DELFT]{Faculty of Mechanical Engineering, Technische Universiteit Delft, The Netherlands}


\begin{abstract}
Designing active-flow-control (AFC) strategies for three-dimensional (3D) bluff bodies is a challenging task with critical industrial implications. In this study we explore the potential of discovering novel control strategies for drag reduction using deep reinforcement learning. We introduce a high-dimensional AFC setup on a 3D cylinder, considering Reynolds numbers ($Re_D$) from $100$ to $400$, which is a range including the transition to 3D wake instabilities. The setup involves multiple zero-net-mass-flux jets positioned on the top and bottom surfaces, aligned into two slots. The method relies on coupling the computational-fluid-dynamics solver with a multi-agent reinforcement-learning (MARL) framework based on the proximal-policy-optimization algorithm. MARL offers several advantages: it exploits local invariance, adaptable control across geometries, facilitates transfer learning and cross-application of agents, and results in a significant training speedup. \rev{For instance, our results demonstrate $16\%$ drag reduction for $Re_D=400$, outperforming classical periodic control, which yields up to $6\%$ reduction.} \revv{A proper-orthogonal-decomposition (POD) analysis at $Re_D=400$ reveals that the DRL control results in a stable wake structure with longer recirculation bubble.} To the authors' knowledge, the present MARL-based framework represents the first time where training is conducted in 3D cylinders. This breakthrough paves the way for conducting AFC on progressively more complex turbulent-flow configurations.

\end{abstract}


\end{frontmatter}

\section*{Introduction} \label{sec:introduccion}

The transportation industry, and the aerospace sector in particular, require new ground-breaking methods to overcome the challenges that they are currently facing, \textit{i.e.} the need to reduce fossil-fuel-related emissions. The implementation of flow-control systems, both passive and active, plays a vital role in the development of more sustainable solutions that can drastically reduce fuel usage, mitigate air and noise pollution, and even improve maneuverability~\cite{choi_review_2008}. The aerodynamic drag in subsonic aircraft is divided mainly into pressure, skin friction (due to viscous stresses), and lift-induced components. Wing-tip effects aside, the two former are the dominant terms.

Control devices utilize aerodynamic principles to manipulate pressure and viscosity, effectively reducing drag. For instance, slats and flaps are control surfaces located on the leading and trailing edges of an airfoil which impact the aircraft operational conditions~\cite{raymer_book_2012}. Modern advancements include winglets~\cite{whitcomb_1976} aimed at mitigating the lift-induced drag or vortex generators~\cite{lin_vortex_2002}, which are used to control boundary-layer separation. Such developments have significantly improved aerodynamic performance. Additionally, a number of alternative approaches like morphing surfaces, spiroids, or blowing devices~\cite{siddiqui_2017, guerrero_2012} are currently being explored. Despite their extensive potential, designing optimal geometries or strategies for these devices has raised significant challenges due to the massive computational resources required to tackle such an intricate interplay between pressure and viscous effects in all flight regimes. Nevertheless, ongoing research efforts and current computational innovations are leading to further advancements toward achieving optimal global control.

In parallel with the recent innovations in flow control, the irruption of machine-learning (ML) techniques has brought tremendous potential to the aeronautics industry, both in terms of studying fundamental problems in fluid mechanics~\cite{vinuesa_transformative_2023, vinuesa_enhancing_2022} and devising completely new strategies for active and passive flow control (AFC and PFC, respectively)~\cite{Le_Clainche_aircraft_2023}. Deep reinforcement learning (DRL) is one of the fastest-growing fields within ML ~\cite{garnier_review_2021} and one of the techniques attracting most interest. Expanding on its success in tabletop games~\cite{alphago_2016}, DRL works well in any system where a controller interacts with an environment to improve a task. That is the case for most AFC cases since DRL interacts with the flow on the fly and receives feedback from it, gaining experience and progressively improving the choice of actions.

AFC setups are complex high-dimensional problems that require substantial computational resources to find the optimal values within the large parametric space of the control system. DRL and neural networks have emerged as valuable tools to make this process feasible, enabling the development of effective control strategies at a reasonable computational cost. The literature on DRL for AFC grows at a fast pace, exhibiting studies on flow control for two-dimensional (2D) cylinders ranging from $Re_D=100$ and $2000$ (where $Re_D$ is the Reynolds number based on inflow velocity $U_{\infty}$ and cylinder diameter $D$) with $17\%$ and $38\%$ drag reduction, respectively~\cite{tang_robust_2020, xu_active_2020, paris_robust_2021, LiZhang_2022, ren_applying_2021, dixia_fan_bluffexperiments_2020, chatzimanolakis2023drag}, aircraft wings~\cite{vinuesa_flow_2022}, fluid-structure interaction~\cite{chen_deep_2023}, turbulent channels~\cite{guastoni_deep_2023}, shape optimization~\cite{yan_2019, viquerat2021direct, keramati_2022}, Rayleigh--Bénard convection~\cite{vignon_effective_2023} or turbulence modelling~\cite{kurz_modelling_2023, novati_2021, beck_toward_2023, bae2022scientific}. Some recent literature demonstrates the possibility of transfer learning from exploration done in 2D cylinders to 3D domains and higher $Re_D$: in Ref.~\cite{Wang_Karniadakis_2023} the wake of a cylinder is controlled by means of two rotating cylinders and in Ref.~\cite{chatzimanolakis2023drag} the control is carried out through multiple jets over the cylinder surface. The present work extends this state-of-the-art in 3D cylinders, considering multiple actuators governed by the novel implementation of a multi-agent reinforcement learning (MARL) framework into a setup based on a distributed-input distributed-output (DIDO) scheme.  In our case, the agent focuses on exploring the underlying 3D physics during training, and the AFC is implemented by multiple independent zero-net-mass-flow (ZNMF) jets placed along the cylinder span and aligned along two slots on the top and bottom surfaces. To the best of the authors' knowledge, this work marks the first time where exploration sessions are directly conducted within 3D cylinders.

As the Reynolds number increases, the flow around a cylinder exhibits different characteristics. Initially, up to approximately $Re_D\approx40$, steady laminar flow prevails, characterized by symmetric counter-rotating vortices in the near wake. Beyond $Re_D\approx190$, laminar vortex shedding emerges, forming the well-known K\'arm\'an vortex street. In the subsequent regimes, between $190<Re_D<260$, the mode-A instability, characterized by dominant spanwise wavelengths of $\lambda_z/D=4$~\cite{williamson_vortex_1996, Barkley_Henderson_1996} is dominant. As $Re_D\approx260$ is surpassed, mode B becomes predominant, and finer three-dimensional features with shorter wavelengths of $\lambda_z/D=1$ are found. Beyond these regimes, the cylinder wake evolves into a more chaotic and turbulent state. 

Discovering flow-control strategies for the flow around a cylinder when the wake transitions from 2D to 3D is challenging. The MARL setup needs to exploit the characteristics of the spanwise structures as the wake becomes three-dimensional to devise effective control approaches. The transition range of for 2D environments has been shown to be suitable, showcasing the generalization ability of deep neural networks~\cite{tang_robust_2020}. However, it has been widely recognized that studying $Re_D>250$ in a 2D context leads to inaccurate predictions of aerodynamic forces. In the present work, the exploration of the 3D context allows tackling possible novel strategies that take advantage of the drag reduction originated by 3D instabilities. 

DRL is based on maximizing a reward $r_t$ provided to an agent interacting continuously with an environment through actions $a_t$. The agent receives information about the environment state at each actuation step through partial observations $s_t$ of the system. This way the agent works on the optimization of a policy $\pi(a_t \lvert s_t)$. A sequence of consecutive actions is denoted as an episode. When a batch $M$ of episodes is finished, the agent updates the neural-network weights to progressively determine a policy that maximizes the expected reward for a given $s_t$. For a detailed understanding of the most recent advances in flow control with MARL we refer to Refs.~\cite{brunton_closed-loop_2015, vignon_recent_2023}. 

\newpage

\section*{Results}\label{sec:results}

This study presents our findings on high-dimensional distributed forcing using multiple jets aligned along the spanwise direction of a 3D infinite cylinder. Training was carried out using multi-agent reinforcement learning (MARL), which demonstrates superior performance compared to conventional single-agent reinforcement learning (SARL) methods. In the subsequent sections, we explore the development of training strategies $\pi(a_t \lvert s_t)$ aimed at achieving high rewards. \rev{We then evaluate the optimal model and compare it with the uncontrolled cases and the periodic control (PC). Note that the uncontrolled converged results across $Re_D$ are obtained after a grid-independence study—more details can be found in Appendix 1. For the study of the periodic control, more details are also provided in Appendix 2, where the tables and heatmaps from the optimization problem are presented. These results will be used for the subsequent analysis. In doing so, we investigate the utilization of the trained DRL agents for exploitation without involving any exploration. Statistical analyses are conducted to elucidate the underlying physical mechanisms responsible for drag reduction, leading to potential energy savings.}

\subsection*{Training} \label{sec:training}

\begin{figure}
    \centering
    \begin{minipage}[b]{0.485\textwidth}
        \centering
        \includegraphics[width=\textwidth,trim = 0.3cm 0.1cm 0.3cm 0.1cm, clip]{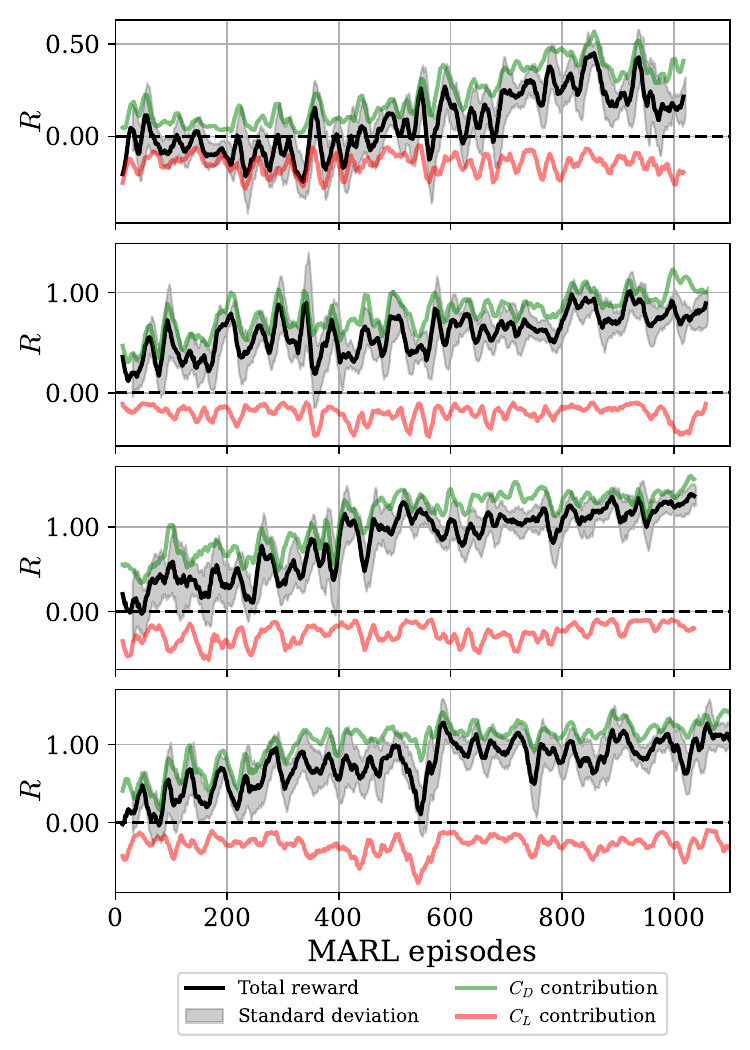}
        \begin{center}(a)\end{center}
    \end{minipage}
    \hfill
        \begin{minipage}[b]{0.49\textwidth}
        \centering
        \includegraphics[width=\textwidth,trim = 0.3cm 0cm 0.23cm 0cm, clip]{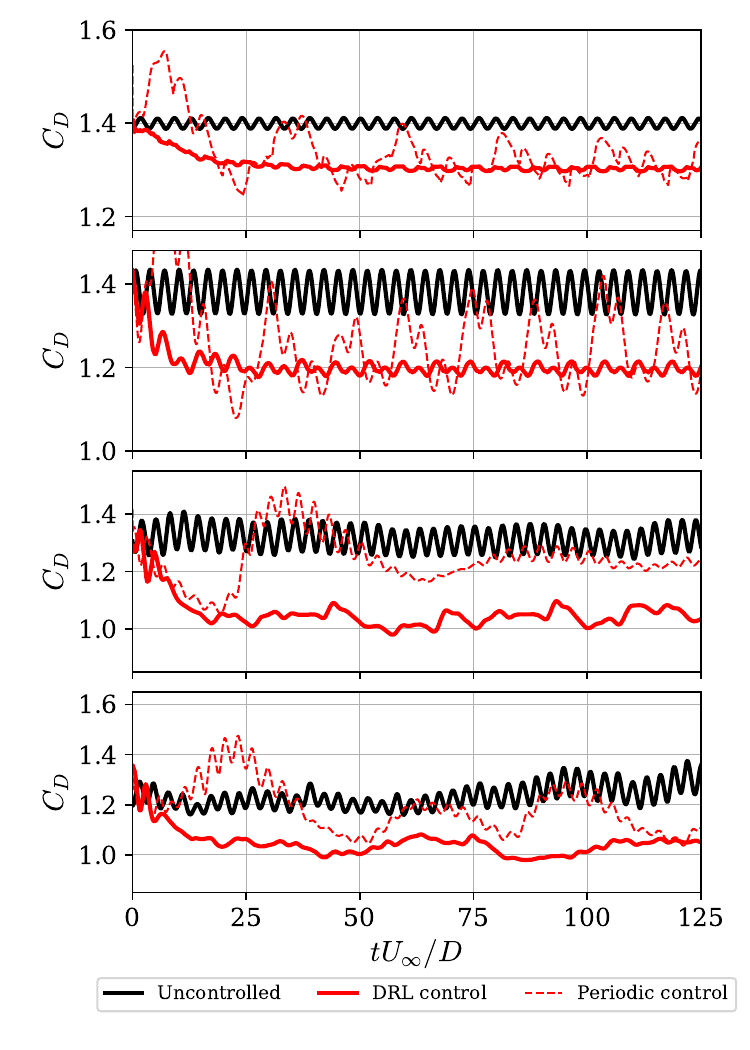}
        \begin{center}(b)\end{center}
    \end{minipage}
    \hfill
  \caption{\rev{\textbf{Reward evolution during exploration episodes and exploitation of the policies.} (a) Final total reward $R$ together with its lift-bias and pure drag-reduction contributions during exploration in training sessions. Signals are smoothed by a moving average of 15 values, and the grey shaded area corresponds to the minimum and maximum rewards over those 15 episodes. (b) Drag-coefficient evolution during exploitation of the model. Comparison between uncontrolled, DRL control and PC. From top to bottom, $Re_D = 100$, $200$, $300$ and $400$.}} 
    \label{fig:fig1}
\end{figure}

To ensure an effective DRL training process, it is crucial to precisely define rewards, penalties, action ranges, and a representative environment state. A fundamental aspect of training is leveraging the physical understanding of the controlled phenomenon to evaluate anticipated reward values and physical control strategies thoughtfully. With these considerations in mind, Figure~\ref{fig:fig1}(a) shows the training curves for the four investigated cases in this study, at Reynolds numbers $Re_D=100$, $200$, $300$, and $400$. Commonly, \rev{sequences of actions $a_t$, states $s_t$, and rewards $r_t$} are referred to as ``environment episodes''. However, in this case, it is more appropriate to call them ``pseudo-environment episodes'' due to the difference between SARL and MARL, where MARL involves multiple pseudo-environments \rev{per environment -- $n_{\rm pseudoenvs} = n_{\rm jets}$ in this case}. Hence, Figure~\ref{fig:fig1}(a) shows all the final rewards from the raw pseudo-environments, together with the pure drag reduction and lift-biased penalization contributions (see the Methods section for more details). As an example, the $Re_D=300$ scenario closely resembles the ideal training condition. This is because the curves exhibit minimal lift bias and result in a total reward that matches the pure drag reduction, stabilizing at $R$ values that are manageable and simple to track. Similar patterns are obtained for the other $Re_D$ cases indicating that the discovered policies are promising for all the cases. Note that we also observe several instances of apparent unlearning such as for $Re_D=400$ at the episode $500$ approximately. This is due to additional exploration of the agent, aimed at increasing the lift asymmetry (note the decrease of the blue line), but quickly returning to exploiting what the agent has identified as a well-performing policy. Once the reward value settles around a certain converged value of $R\approx 1.0$, the training process is concluded. \rev{The moment of stopping the training is always associated with a high-risk of both unlearning and having much more computational burden than expected. The best way to assess when training is concluded is to exploit the model with no exploration at different stages and observe any drop or improvement in performance.} Then, we proceed to the assessment of the learned policies in deterministic mode, which is discussed next.

In terms of computational cost, training is the most significant part. On average, each training session requires about $1200$ MARL episodes, which is equivalent to running $120$ numerical simulations for the entire domain. All exploration sessions were conducted on the Dardel high-performance computer in the PDC Center at KTH Royal Institute of Technology. The sessions run on 8 nodes simultaneously, each running one numerical simulation comprising 10 simultaneous pseudo-environments. Hence, 80 pseudo-environments in total. Each node has two AMD EPYC™ Zen2 2.25 GHz 64-core processors with 512 GB memory. With each batch of 8 simulations taking ideally five hours in this particular architecture, it requires less than four days of continuous operation. Moreover, as the Reynolds number ($Re_D$) increases, the computational cost increases significantly.


\subsection*{Exploitation of the models} \label{sec:exploit}

At this point, the agent policies are evaluated without any exploration. As a result, the agent calculates the most likely value of the action $a_t$ within its learned probability distribution, aiming to maximize the expected reward. In Figure~\ref{fig:fig1}(b) we show the temporal evolution of the DRL control and the PC. The DRL-based control exhibits a clear two-phase process that starts with a short transient period followed by the stationary control policy. We included the transition phase between $t=0$ to around $20$ convective time units. Note that the vortex-shedding period is $T_k=1/St\approx1/0.2 = 5$, where $St=f D/U_{\infty}$ is the Strouhal number and $f$ is the vortex-shedding frequency. These results imply that it takes less than $4T_k$ to reach the stationary behavior. The DRL-based control exhibits a first suction/ejection overshoot which destabilizes the wake, and then it proceeds to re-stabilize it in a second phase. During the latter, the jet mass flux exhibits lower values which barely reach $75\%$ of those in the transient overshoots. 

\rev{The time window shown in Figure~\ref{fig:fig1}(b) is limited to 125 time units to better highlight the two-phase behavior of the DRL control. However, the actual simulations were run for at least 200 time units to ensure temporal convergence, with statistical quantities computed over the entire duration, excluding the initial transients.}

\begin{figure}[H]
    \centering
    \begin{minipage}{\textwidth}
        \centering
        \begin{subfigure}{\textwidth}
            \centering
            \includegraphics[width=\textwidth]{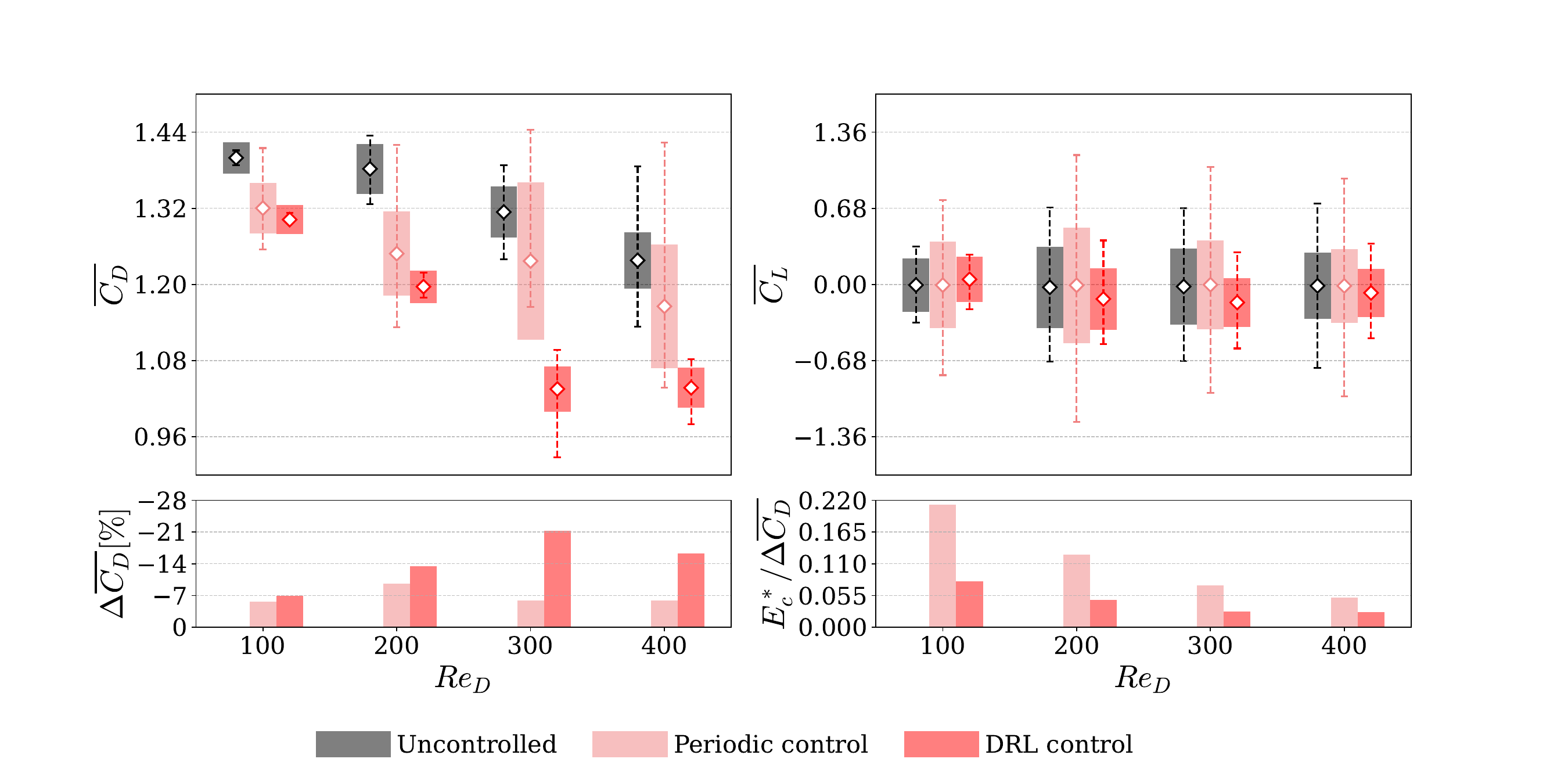}
            \subcaption*{(a)}
            \vspace{4pt} 
            \label{fig:cd_cl_rms}
        \end{subfigure}

        \begin{subfigure}{\textwidth}
        \centering
        \includegraphics[width=0.85\textwidth, trim=4cm 4cm 4cm 4cm]{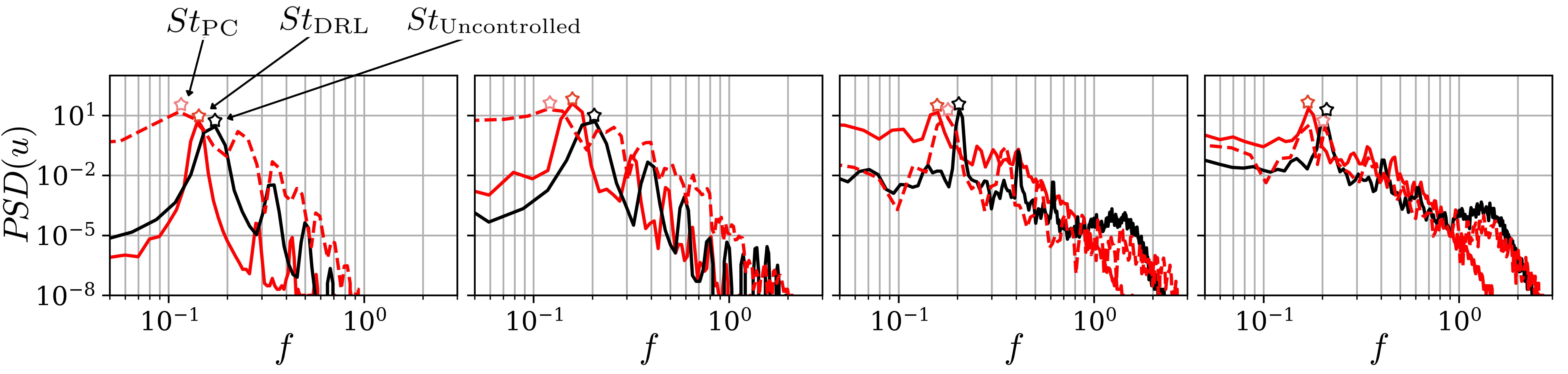}
        \subcaption*{(b)}
        \vspace{5pt} 
        \label{fig:spectra}
        \end{subfigure}
        
        \centering
        \resizebox{0.95\textwidth}{!}{
        \begin{tabular}{c||cc|cccc|cccc}
            & \textbf{Williamson (1992) \cite{williamson_vortex_1996}} & \textbf{Uncontrolled} & \multicolumn{4}{c|}{\textbf{Periodic control}} & \multicolumn{4}{c}{\textbf{DRL control}}    \\
        $Re_D$       & \multicolumn{2}{c|}{$St$}                & $Q_{\rm{max}}$ & $Q_{\rm{RMS}}$ & $f_c$            & $St$         & $Q_{\rm{max}}$ & $Q_{\rm{RMS}}$ & $f_c$          & $St$       \\
        \hline    
        100    & 0.165 & 0.170      & 0.053  & 0.037   & 0.115    & 0.113     & 0.016  & $8.4 \times 10^{-3}$         & 0.144         &  0.139      \\
        200    & 0.185 (Mode A) & 0.186      & 0.053  & 0.037   & 0.130    & 0.117    & 0.031  & $9.5\times 10^{-3}$         & 0.160         & 0.157        \\
        300    & 0.203 (Mode B) & 0.206      & 0.018  & 0.013   & 0.175    & 0.177     & 0.031  & $7.6\times 10^{-3}$        & 0.158         & 0.159        \\
        400    & 0.205 (Mode B) & 0.202      & 0.012  & 0.008   & 0.172    & 0.194     & 0.025  & $5.7\times 10^{-3}$    & 0.171        & 0.171       
        \end{tabular}}
        \subcaption*{(c)}
        \label{tab:Table}
    \end{minipage}
    
    \caption{\rev{\textbf{Summary of aerodynamic forces, wake spectra and mass-flow rate characteristics for the various controls.} (a) Mean drag ($C_D$) and lift ($C_L$) coefficients (white diamonds), RMS fluctuations (thick bars), and max-min range (dashed intervals). Percentage drag reduction $\Delta \overline{C_D}$ and cost metric $E_{c}^*/\Delta \overline{C_D}$ (lower is better) from Equation (\ref{eq:Ec}). (b) Power-spectral density of streamwise velocity $u$ at $x/D=10.5$ for uncontrolled (black), PC (red dashed), and DRL-based control (red). From left to right: $Re_D=100,200,300$ and $400$. (c) Summary of control strategies: Strouhal number $St$ (compared with literature), max and RMS mass-flow rates ($Q_{\rm max}$, $Q_{\rm RMS}$), and control frequency $f_c$ (the optimum found for the sinusoidal signal of PC, and the peak from DRL actuators signal).}}
    \label{fig:fig2}
\end{figure}

This control strategy persists until control stabilizes into stationary behavior, which is monitored by assessing mean quantities and fluctuations in aerodynamic forces. The averaged drag-reduction results for all $Re_D$ are reported in Figure~\ref{fig:fig2}(a). It is important to note that all the cases lead to effective drag-reduction rates. The overall performance is much better than what can be obtained with the classical PC strategies. \rev{In summary, the percentage changes in the mean drag coefficient are $\Delta \overline{C_{D}}\vert_{\rm DRL} = -7\%$, $-13.4\%$, $-21.2\%$, $-16.2\%$, compared to $\Delta \overline{C_{D}}\vert_{\rm PC} = -5.7\%$, $-9.7\%$, $-5.9\%$, $-5.9\%$ for $Re_D = 100$, 200, 300, and 400, respectively. We acknowledge that there is also a notable reduction in lift oscillations as a consequence of the drag reduction mechanisms employed by the DRL agents, which were not directly addressed in the reward $r_t$ function. This trade-off is an important aspect and presents an opportunity for future approaches that may consider maximizing lift RMS or other metrics depending on the application.} All quantities of interest are averaged in time by considering an interval of at least $20T_k$, {\it i.e.} over 100 time units, excluding the transients obtained after applying the control. The root-mean-square of the fluctuations, \rev{$\phi_{\rm RMS} = \sqrt{(1/n)\sum_{i=1}^{n}(\phi_i-\overline{\phi})^2}$}, minimum and maximum values provide deeper insights into the mentioned robustness. While the mean values alone may suggest a good performance of the PC, the merits of the control should not be assessed solely based on this quantity. When considering an optimal control strategy, the preferred choice typically involves selecting a control with minimal variability and few extreme values, which are characteristics exhibited as evidenced by the DRL-based control.

We also study the ratio between the total fluid mass intercepted by the frontal area of the cylinder $E_{\infty}$, \rev{where $Q_\infty=DU_\infty$}, and the total mass used by the actuators $E_c$. Based on the definitions used in Ref.~\cite{chatzimanolakis2023drag}, we propose the following expression for the ratio $E_c^*$: 

\begin{equation} \label{eq:Ec}
E_{c}^* = \frac{E_c}{E_\infty} = \frac{L_{{\rm jet}}}{(t_2 - t_1) Q_\infty L_z} \bigintss_{t_1}^{t_2} \sum_{i=1}^{n_{\rm{jets}}} \lvert Q_i(t) \lvert \ \rm{d}t,
\end{equation}

\noindent where $t_1$ and $t_2$ define the start and end of our time interval for evaluating the control. \rev{Note that while this evaluation of mass consumption is based on a numerical approach with modified Dirichlet boundary conditions, the actual mass cost would depend on the specifics of the experimental setup, including the type of actuators, such as membranes or reservoirs, and the jet configurations used. In the absence of experimental data for these components, we adopt this numerical approach. However, the present results indicate that the mass cost associated with DRL-based control remains minimal in comparison to the significant drag reduction achieved. This emphasizes the efficiency of DRL-based control strategies, as illustrated by the results in Figure~\ref{fig:fig2}, where the $E_{c}^*/\Delta \overline{C_D}$ ratio demonstrates that DRL requires only half the mass of classical control methods to achieve a comparable drag reduction.}

We provide additional physical insight by assessing the power-spectral density (PSD) of the streamwise velocity, shown in Figure~\ref{fig:fig2}(b). This figure illustrates how the change of frequency impacts the wake topology after applying the various control strategies. In particular, both the DRL-based control and PC cases exhibit a reduction in $St$. Further insight into the various control strategies is provided in Figure~\ref{fig:fig2}(c), where several characteristic variables of the various controls are shown. The first important observation is the fact that the root-mean-square (RMS) of the jet mass-flow rate (computed by averaging in time and the spanwise direction) is one order of magnitude lower in the DRL than in the PC. This indicates that the DRL-based control strategies lead to more stable and robust configurations, avoiding large peak-to-peak variations in the actuation. Also, note that we show the control frequencies $f_c$ for the PC (where we report the optimal control frequency) and the DRL (where we report the dominant frequency). Although the $f_c$ values are not dramatically different in the PC and DRL cases, the latter exhibit significantly more complex control laws than the former. It is worth noting that we conducted a parametric study to assess the optimal values of $f_c$ and $Q$ for classical PC, as discussed in the \rev{Methods section and referred in Appendix 2.}

\begin{figure}[h]
    \centering
    \includegraphics[width=\textwidth]{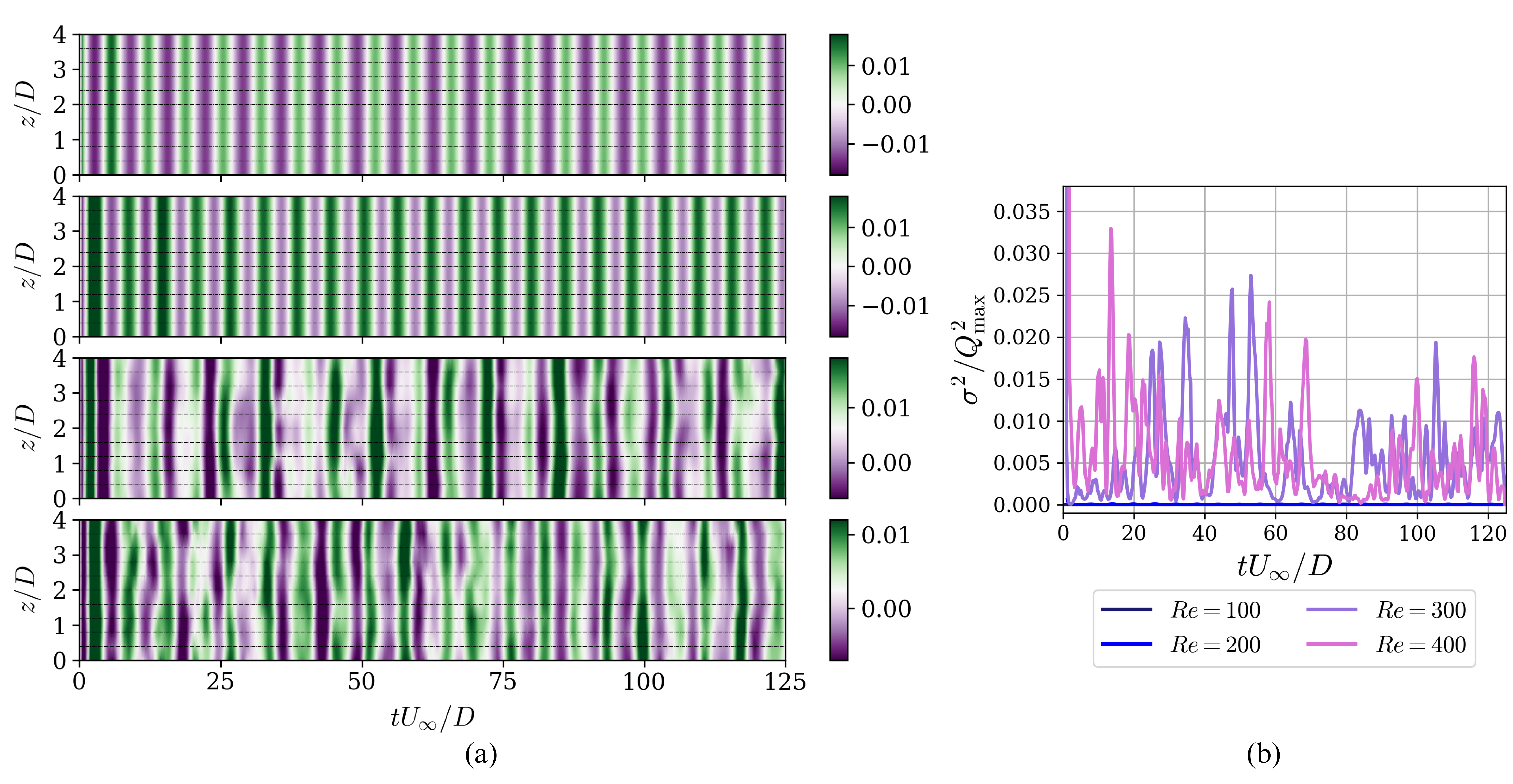}
  \caption{\textbf{Evolution of the mass-flow rate associated with the jets in time and in the spanwise direction.} (a) Mass-flow rate per unit width $Q$ as a function of time for all jets individually, showing also their spanwise distribution for the DRL cases. From top to bottom: $Re_D = 100$, $200$, $300$ and $400$. (b) Evolution in time of the variance of the mass-flow rate computed in $z$, $\sigma^2(t) = (1/n_{\rm{jets}}) \sum_{i=1}^{n_{\rm{jets}}} (Q_i(t)- \overline{Q}(t))^2$, for the different Reynolds numbers under study. Note that $\sigma^2$ is normalized by the squared peak $Q$ values from each case, and $\sigma^2=0$ is obtained for $Re_D=100$ and $200$.} 
  \label{fig:fig3_mfr}
\end{figure}

Figure~\ref{fig:fig3_mfr} demonstrates the main advantage of a MARL implementation: the control policy can act locally, exploiting wake vortical structures and distributing the jet flow in the spanwise direction to minimize overall drag. In Figure~\ref{fig:fig3_mfr}(a), we show the temporal and spanwise evolution of the mass-flow rate per unit length from the jets under the DRL-based control strategy. The agent utilizes less than 10\% of the maximum possible value, as discussed in the Methods section. For $Re_D \leq 200$, the mass flow is uniformly distributed in the spanwise direction, while beyond this Reynolds number, the control begins to introduce spanwise variations. This is supported by the results in Figure~\ref{fig:fig3_mfr}(b), which shows the instantaneous variance of the mass flow rate in $z$ over time for the various cases, denoted as $\sigma^2$. As mentioned in the Introduction, for $Re_D \geq 250$, the wake displays three-dimensional features, which are exploited by the DRL control to maximize drag-reduction rates.

During the exploration stage, for $Re_D=100$ and $200$, the agent did not find any spanwise-varying strategies that improved performance, suggesting that the wake is two-dimensional in these cases, favoring spanwise-uniform control strategies. In contrast, for $Re_D=300$ and $400$, flow patterns associated with transitional $Re_D$ emerge, including spanwise structures of approximately one cylinder diameter ($\lambda_z/D=1$), related to mode-B instabilities. Additionally, a basic SARL approach does not effectively exploit these local spanwise scales, while the MARL setup enables the use of these structures, as shown by the non-zero variance in the control in $z$.

\rev{Although it might seem plausible that a single-agent SARL approach could achieve spanwise-uniform strategies due to its simplicity, computational limitations make this approach unfeasible. Specifically, SARL faces the curse of dimensionality, requiring exponentially more trajectories to explore the state space thoroughly. This results in a significantly higher computational cost compared to the multi-agent DRL approach, which is more efficient in navigating complex control strategies.}

In Figure~\ref{fig:fig4_cartoon} we illustrate how the flow topology is influenced by the various drag-reduction strategies, on three representative phases: uncontrolled, transient, and stabilized control. The flow visualizations indicate that the control strategies based on DRL aim to enhance the spacing between successive vortical structures, resulting in a reduction of the vortex-shedding period $T_k$. Hence, mode-B instabilities are diminished when the control is applied, and the intensity of the vortex shedding is attenuated. These changes lead to a more organized wake structure, resembling the characteristic two-dimensional laminar wake. Figures~1(b) and 2(a) corroborate these findings, illustrating diminished oscillations during the controlled phase.

\rev{The observed reduction in unsteady structures within the recirculation bubble, or the increased absence of vortical activity in this region, can be associated with changes in the drag-reduction mechanisms. This interpretation is supported by statistical data presented later in this section in Figure~\ref{fig:fig5_fields}, which provides a more robust perspective on the flow dynamics.}


\begin{figure}[H]
\begin{subfigure}{\textwidth}
            \centering
            \includegraphics[width=0.85\textwidth]{figs/NATURE_fig4_cartoon_a.pdf}
            \caption{}
        \end{subfigure}
        
        \vspace{0.2cm}
        \begin{subfigure}{\textwidth}
        \centering
        \includegraphics[width=0.85\textwidth]{figs/NATURE_fig4_cartoon_b.pdf}
        \caption{}
        \end{subfigure}
\caption{\rev{\textbf{Visualizations of the flow coherent structures during DRL-based strategy exploitation.} Temporal evolution of the flow at (a) $Re_D=300$ and (b) $Re_D=400$, from uncontrolled state to a stable DRL control. (Top) Non-dimensional mass-flow rate per unit width $Q$, lift coefficient $C_L$ and drag coefficient $C_D$ as a function of time; (bottom) snapshots showing vortical structures identified with the $\lambda_2$ criterion~\cite{Jeong_Hussain_1995}, where the isosurface $\lambda_2 D^2/U_{\infty}^2=-0.5$ is shown for uncontrolled, transient, and DRL stabilized control states. The colors framing the flow visualizations correspond to the instants indicated in the temporal evolution of the relevant flow quantities.}} 
\label{fig:fig4_cartoon}
\end{figure}

Studying flow statistics offers deeper insight into the mechanisms employed by the DRL agent to discover flow-control strategies, particularly when analyzing the mean flow and the Reynolds stresses. To compute the latter, the Reynolds decomposition is used to decompose the flow variables ($u$) into time-averaged mean ($\overline{u}$) and fluctuating ($u'$) components, $u = \overline{u} + u'$. In Figures~\ref{fig:fig5_fields}(a), we observe the impact of the DRL-based control: the wake nearly doubles the recirculation-bubble length, delaying the wake-stagnation point by approximately one diameter in the streamwise direction. Instead of showing all cases, we only present $Re_D=400$ as a representative case. Figures~\ref{fig:fig5_fields}(b) and \ref{fig:fig5_fields}(c) show how the wake also changes noticeably, being slightly wider but decaying much faster as we move downstream. Note that peaks in $v$ fluctuations follow the same pattern, meaning that the counter-rotating vortices occur further as well. The pressure coefficient $C_p = 2(P - P_{\infty})/(\rho U_{\infty}^2)$ around the cylinder is shown in Figure~\ref{fig:fig5_fields}(d) for the DRL-controlled case and the uncontrolled one. The back pressure increases by $\Delta \overline{C_{pb}} \approx 0.4$, which is directly related to the drag reduction mechanism. 

The Reynolds stresses are presented in Figure~\ref{fig:fig5_fields}(e), which shows that the peaks move downwards in the streamwise direction after applying the control, with only small changes in the vertical location. Additional analysis is provided in Figure~\ref{fig:fig5_fields}(f), where the DRL-based control generally leads to the reduction of the peak magnitude in almost all the fluctuating quantities. In this case, all $Re_D$ values are presented to elucidate that the same behavior occurs within this regime range. When considering fluctuations in the spanwise direction $w'$, we notice a distinct pattern: an increase occurs at $Re_D = 300$, while a decrease is observed at higher $Re_D$ values.

\begin{figure}[H]
    \centering
\begin{subfigure}{\textwidth}
    \centering
    \includegraphics[width=0.99\textwidth,trim = 1.75cm 1cm 1.75cm 0cm, clip]{figs/NATURE_fig5_restre_avg_review_a.pdf}
    \caption{}
\end{subfigure}

\begin{subfigure}{0.33\textwidth}
    \centering
    \includegraphics[width=\textwidth,trim = 0.32cm 0.22cm 1cm 0.1cm, clip]{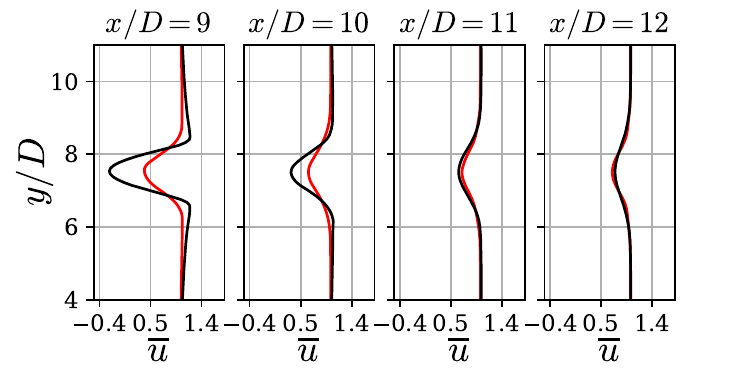}
    \caption{}
\end{subfigure}
\begin{subfigure}{0.33\textwidth}
    \centering
    \includegraphics[width=\textwidth,trim = 0.32cm 0.22cm 1cm 0.1cm, clip]{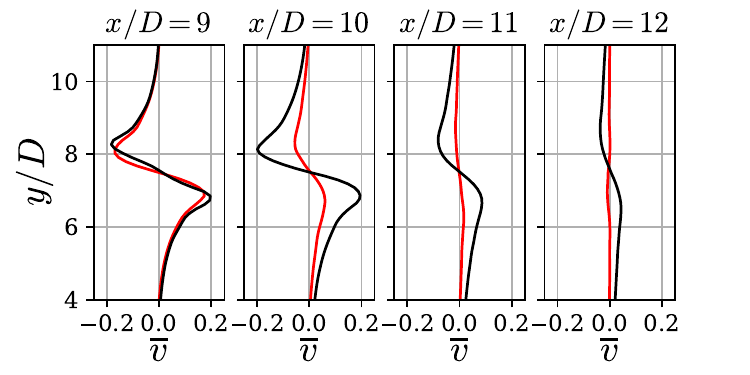}
    \caption{}
\end{subfigure}
\begin{subfigure}{0.30\textwidth}
    \centering
    \includegraphics[width=\textwidth,trim = 0.32cm 0.22cm 2cm 0.1cm, clip]{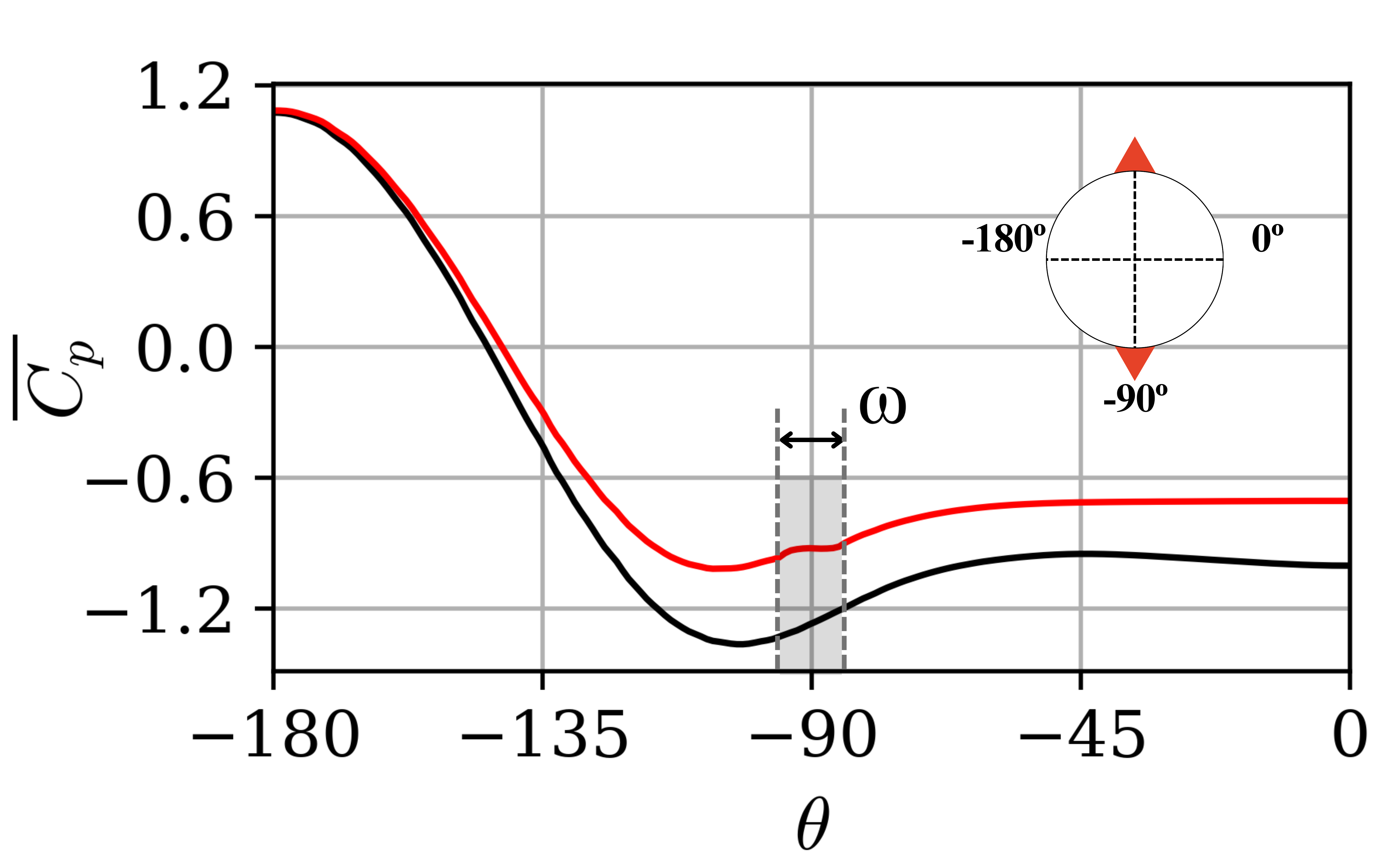}
    \caption{}
\end{subfigure}

\begin{subfigure}{\textwidth}
    \centering
    \includegraphics[width=0.99\textwidth,trim =3.25cm 0.28cm 2.5cm 0cm, clip]{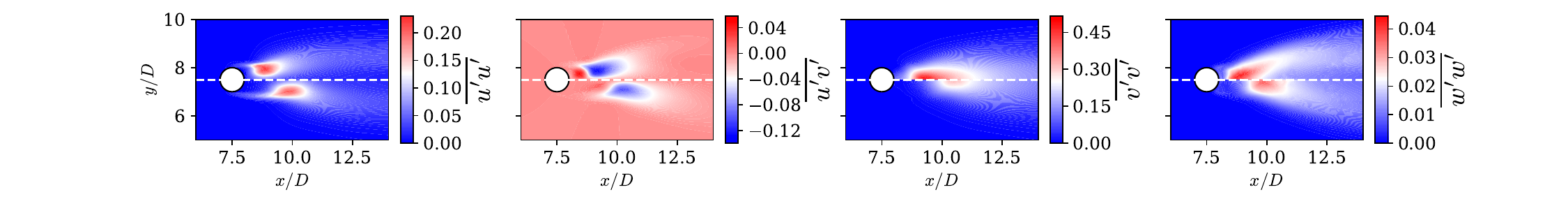}
    \caption{}
\end{subfigure}

\begin{subfigure}{\textwidth}
    \centering
    \includegraphics[width=0.99\textwidth,trim =0cm 0cm 0cm 0.5cm, clip]{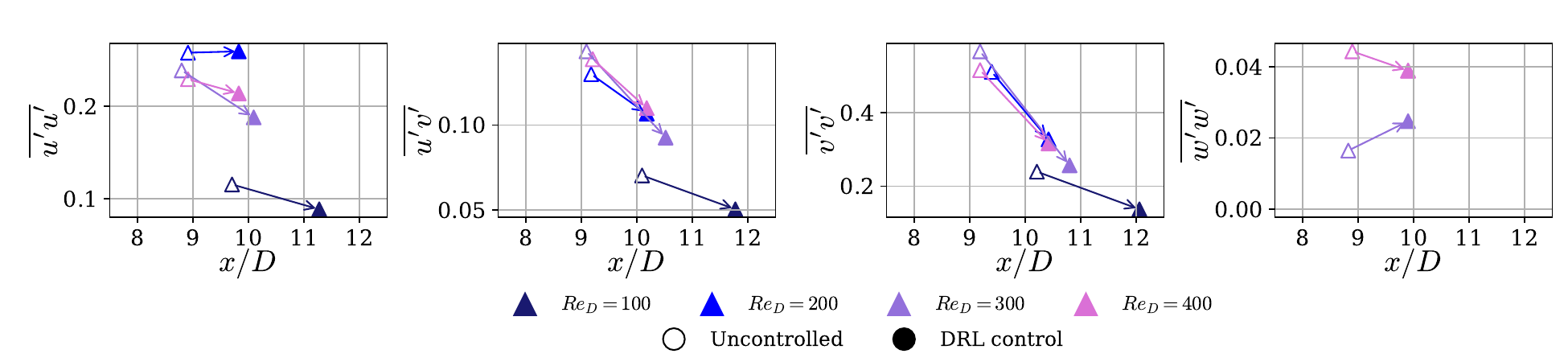}
    \caption{}
\end{subfigure}

\caption{\rev{\textbf{Mean flow and Reynolds stresses for the DRL-controlled cases.} (a) Mean velocities and pressure fields for $Re_D=400$, where (half top) is uncontrolled and (half bottom) is DRL-controlled flow. Yellow lines denote the regions where $\overline{u}=0$ which indicate the wake-stagnation points also annotated with their $x/D$ location. Mean wake profiles of (b) $\overline{u}$ and (c) $\overline{v}$, as well as (d) mean pressure distribution $\overline{P}$ on the cylinder, respectively. We show (black) uncontrolled and (red) controlled cases. \rev{Note that the gray shaded area in (d) is the arc covered by the jet.} (e) Reynolds stresses, from left to right, $\overline{u'u'}, \overline{u'v'}, \overline{v'v'}$ and $\overline{w'w'}$, where (half top) is uncontrolled and (half bottom) is DRL-controlled flow. (f) Peak values for Reynolds stresses and their $x/D$ locations for all the Reynolds numbers under study. Note that $\overline{w'w'}$ for $Re_D=100$ and $200$ is not displayed because it is zero.}}
\label{fig:fig5_fields}
\end{figure}

\begin{figure}[H]
    \centering
\begin{subfigure}{\textwidth}
    \centering
    \includegraphics[width=0.99\textwidth]{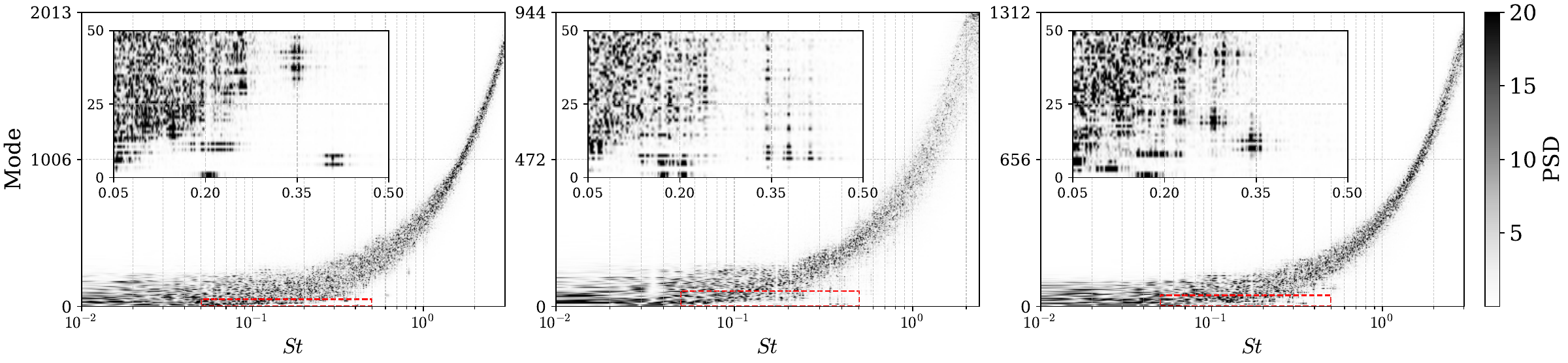}
    \caption{}
\end{subfigure}

\begin{subfigure}{\textwidth}
    \centering
    \includegraphics[width=0.99\textwidth]{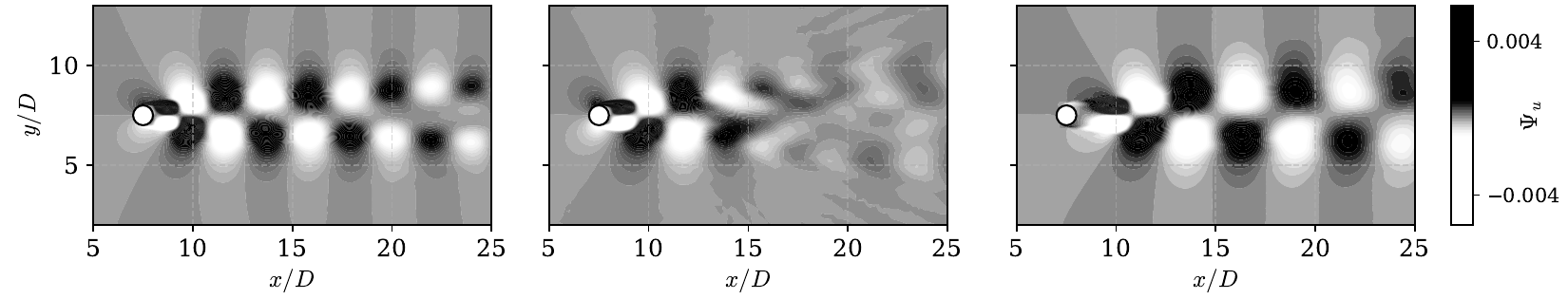}
    \caption{}
\end{subfigure}
\begin{subfigure}{\textwidth}
    \centering
    \includegraphics[width=0.95\textwidth, trim = {0.1cm 0 0 0}, clip]{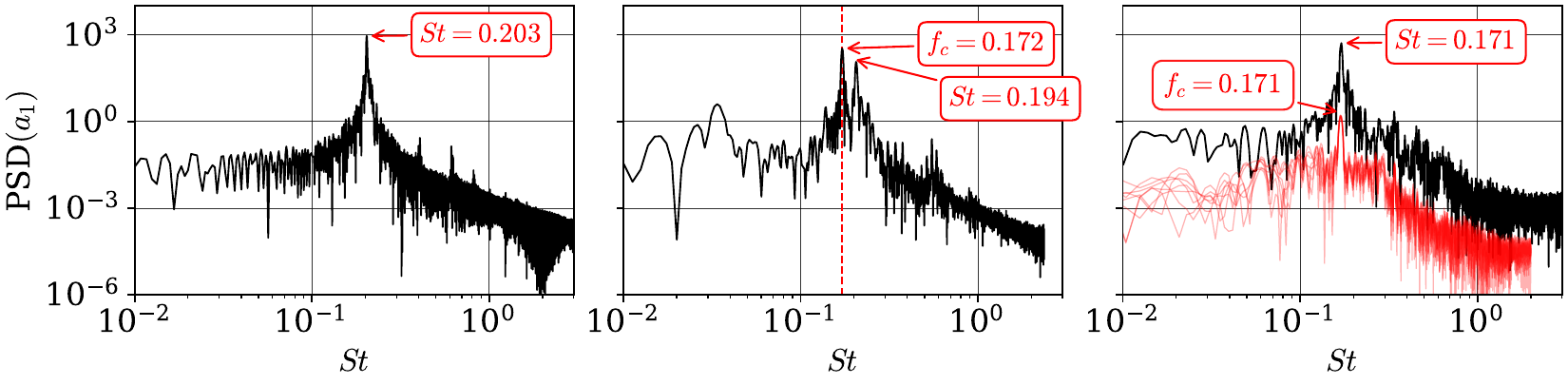}
    \caption{}
\end{subfigure}

\caption{\revv{\textbf{Proper-orthogonal-decomposition (POD) results for streamwise velocity $u$ at $Re_D=400$ comparing, from left to right: uncontrolled, PC, and DRL-controlled case.} (a) Heatmap of the full spectrum for the POD modes and insert on top-left corner zooming into the first 50 modes within the dominant $St$. (b) Streamwise velocity for the first POD mode in the $xy$-plane (homogeneous in the spanwise direction). (c) Power-spectral densities corresponding to the first POD mode, highlighting the shedding Strouhal number ($St$) and the characteristic frequency ($f_c$), which appears as a double peak in the PC case. Note that for PC, a dashed red vertical line \( f_c \lvert_{\rm PC} \) is shown, while for DRL, the spectra of all \( Q_i \) are displayed in red with an offset to also highlight \( f_c \lvert_{\rm DRL} \).}}
\label{fig:fig6_POD}
\end{figure}

\revv{Additionally, we performed a proper-orthogonal-decomposition (POD) analysis \citep{lumley1967_POD} of the uncontrolled, PC and DRL-controlled cases at $Re_D=400$ to better understand their flow physics. The PODs were performed with the pyLOM package \citep{EIXIMENO2025109459} on the streamwise velocity and pressure fields over the last 40$T_k$, with a sampling rate of about 25 snapshots per vortex shedding.

Inspection of the frequency content of the temporal coefficients shown in Figure~\ref{fig:fig6_POD}(a) reveals that the DRL control acts selectively and less invasively. PC acts on a single frequency, resulting in a perturbation of the flow frequencies that can be seen in a more diffuse spectrum with a wider tail, thus indicating a stronger perturbation of the flow. In contrast, DRL control identifies and selects a wider range of frequencies to act on, resulting in a spectrum that is closer to that of the uncontrolled flow. As, a result, the flow exhibits a narrower range of frequencies yielding a less perturbed state.

Moreover, the streamwise component of the first POD mode is shown in Figure~\ref{fig:fig6_POD}(b) and the corresponding power-spectral densities are shown in Figure~\ref{fig:fig6_POD}(c). This mode contains a significant contribution of the vortex-shedding for the uncontrolled flow case. In the PC there is, in addition to the identified vortex shedding at $St\lvert_{\rm PC}=0.194$, as reported in Figure~\ref{fig:fig2}(c), the contribution of the actuation frequency at $f_c\lvert_{\rm PC}=0.171$. For the DRL-controlled flow we observe that the first mode closely resembles the control strategy found by the agent. Thus, DRL control is able to act on a wide range of frequencies, while in PC only acts at a single frequency. This effectively alters the vortex shedding frequency of the DRL-controlled flow to $St\lvert_{\rm{DRL}}=0.171$. In fact, DRL control actively works to modify the vortex shedding frequency from the uncontrolled state, $St\lvert_{\rm Uncontrolled}=0.2$. As a consequence, there are significant differences in the flow characteristics. In PC the resulting modes become highly disturbed on the far-wake, while the near-wake remains mostly unaltered, exhibiting a strong similarity compared with the uncontrolled case. Furthermore, for the PC, it is observed that the double peak of $f_c \lvert_{\rm PC}$ and $St\lvert_{\rm PC}$ coexist, indicating that the PC is not successfully modifying the shedding in the same way as the DRL-controlled system does.

On the other hand, in the DRL-controlled flow the structures exhibit a more elongated wake structure and a larger recirculation bubble (measured from the end of the cylinder) of approximately $L_r/D = 2$, in contrast to the uncontrolled and PC recirculation bubbles of approximately $L_r/D = 0.8$ and $L_r/D = 0.9$, respectively. Subsequent modes exhibit more complex structures related to sub-harmonics and turbulent transition. The DRL-controlled flow is associated with modes that break the flow symmetries, even in the span-wise direction, suggesting a transition to three-dimensional flow at more energetic modes.}


\section*{Discussion and conclusions}\label{sec:conclusion}

In this study a multi-agent reinforcement-learning (MARL) framework is coupled with a numerical solver to discover effective drag-reduction strategies by controlling multiple jets placed along the span of three-dimensional cylinders. We study cases at $Re_D=100$, $200$, $300$, and $400$, where wake transition from 2D to 3D is observed. All DRL-based control policies outperform the classical periodic control in this $Re_D$ range. This is characterized by the emergence of spanwise instabilities, which the DRL agent can exploit to discover effective drag-reduction strategies. This is achieved by taking advantage of exploiting the underlying physics within pseudo-environments and optimizing the global problem involving multiple interactions in parallel. One of the main advantages of employing MARL is the capability to deploy trained agents across various cylinder lengths and numbers of actuators while ensuring consistency in the spanwise width of the jets ($L_{\rm{jet}}$) and their corresponding pressure values as observation states ($s_t$). Note that the training focuses on symmetries and invariant structures. This would not be possible with SARL, which is restricted to a certain number of actuators (and also the corresponding algorithm limitations). MARL allows for cheaper training sessions in smaller and simplified computational domains, thereby speeding up the process, which is required to perform flow control in high-fidelity simulations.

These findings highlight the effectiveness of the DRL approach, which can discover flow-control strategies more sophisticated than those obtained with the classical periodic control, spanning wide ranges of frequencies and tackling different flow features in the wake. DRL-based control achieves a remarkable performance, reducing drag by $21\%$ and $16.5\%$ for $Re_D=300$ and $400$ respectively, outperforming PC strategies which only achieve around $6\%$ reduction for both $Re$. \revv{A POD of the DRL-controlled flow revealed the underlying physics of the actuation at $Re_D=400$, resulting in a stable wake structure with a longer recirculation bubble.}

\rev{In fact, addressing the curiosity of whether these DRL results could be simplified or reduced to a basic control—one resembling a periodic control, but already knowing the dominant frequencies from the DRL that yield such good performance and range of $Q$—was investigated. The idea was perhaps to find strategies that do not require solving such a high-dimensional problem. However, the results reported in the second part of Appendix 2 indicate that achieving drag reduction is not feasible without considering the full spectrum of both temporal and spatial frequencies utilized by MARL agents. The results show that alternative approaches do not even achieve a quarter of the performance on average. This wider spectrum of closed-loop strategies is essential for addressing the three-dimensionalities and transitional stages required to shift from an uncontrolled to a stabilized controlled state.}

Furthermore, the results presented here represent the first training conducted in 3D cylinders using a MARL implementation. \rev{This study sets a new benchmark for the DRL community, potentially inspiring its application to more complex turbulent scenarios at higher Reynolds numbers and to other Distributed Input-Distributed Output (DIDO) frameworks. Additionally, future research could improve training efficiency by implementing the approach in Ref.~\cite{jeon2024_invariants}, which explores the use of group invariants and positional encoding.}

\section*{Methods}\label{sec:method}

\subsection*{Problem configuration and numerical setup} \label{sec:setup}

\begin{figure}[H]
    \centering
    \begin{subfigure}{\textwidth}
        \centering
        \includegraphics[width=0.7\textwidth]{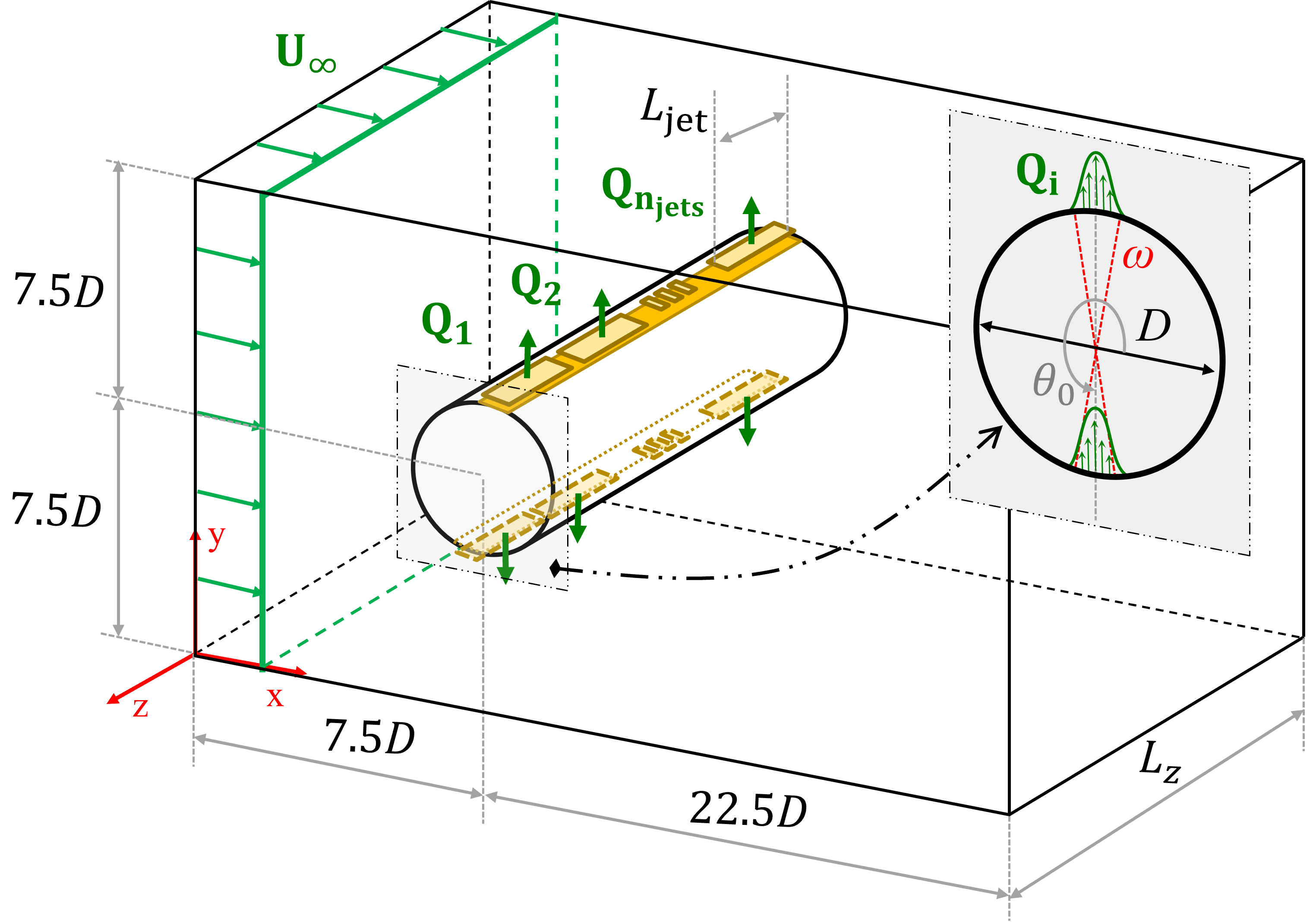}
        \caption{}
    \end{subfigure}
    
    \vspace{0.1cm} 
    \begin{subfigure}{0.7\textwidth}
        \centering
        \includegraphics[width=\textwidth]{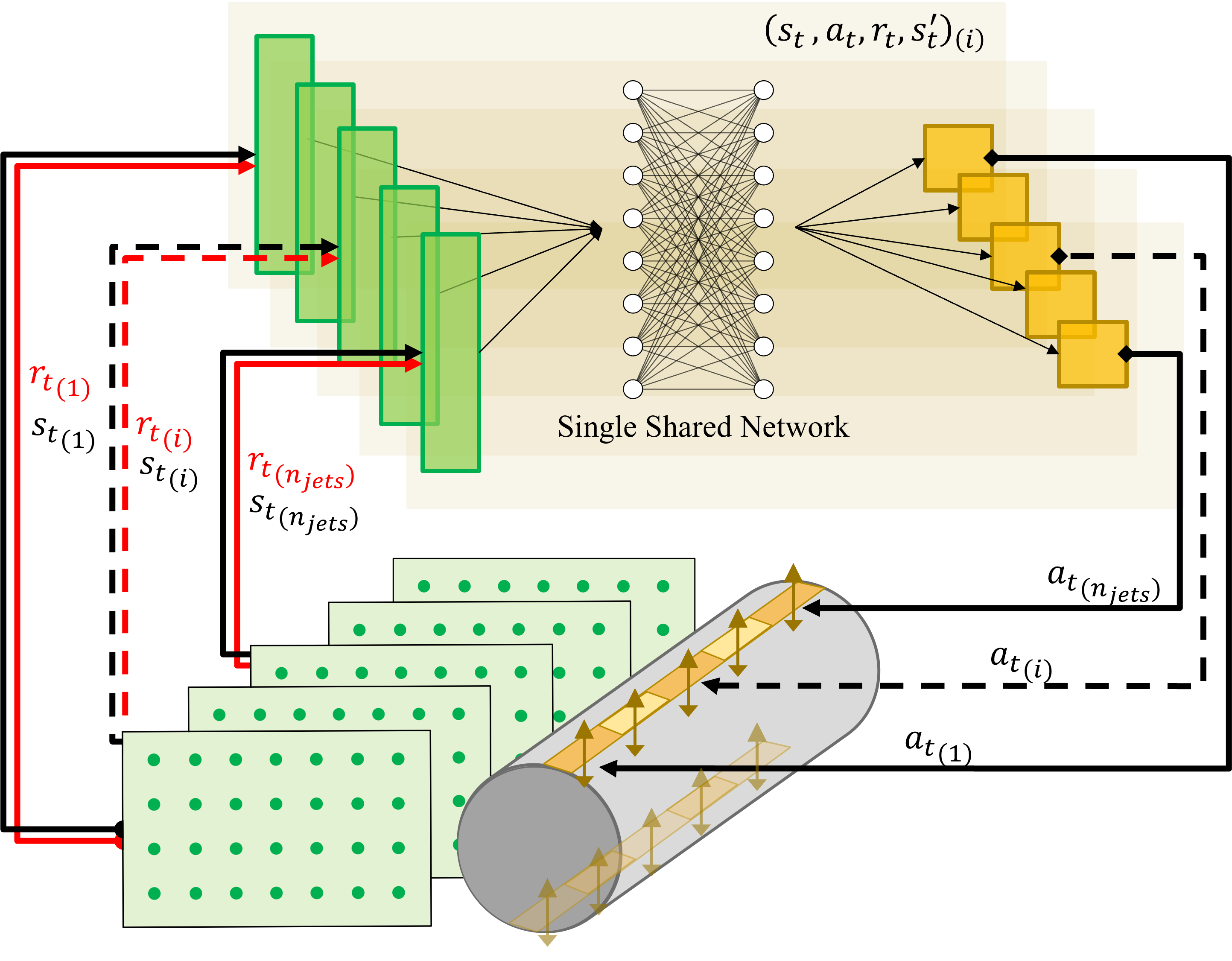}
        \caption{}
    \end{subfigure}

    \caption{
        \rev{\textbf{Conceptual visualization of the computational domain, the MARL framework, and a summary of parameters.} (a) Schematic representation of the computational domain with cylinder diameter $D$ as the reference length. Here $\omega$ is the jet width and $\theta_0$ is the angular location of each jet. 
        In green, we show the velocity condition for the inlet $U_{\infty}$ and the sinusoidal profile in the jet azimuthal direction. (b) MARL framework applied to a three-dimensional cylinder equipped with distributed input distributed output (DIDO). Note that both (a) and (b) are not to scale.}
    }
    \label{fig:fig6_sketch}
\end{figure}

The present study consists of a 3D cylinder exposed to a constant inflow in the streamwise direction. All domain lengths are non-dimensional, using the cylinder diameter $D$ as the reference length. The geometry under consideration is presented in Figure~\ref{fig:fig6_sketch}(a). The computational domain has a streamwise length of $L_x/D=30$, a height of $L_y/D=15$, and a spanwise length of $L_z/D=4$. The cylinder is centered at $(x/D,y/D)=(7.5,7.5)$. Since the cylinder is considered to be infinitely long in the spanwise direction, we use periodic boundary conditions in $z$. Furthermore, we use a Dirichlet condition with a constant velocity $U_{\infty}$ at the inlet. The top, bottom and outflow surfaces are outlets with imposed zero velocity gradient and constant pressure. The cylinder surfaces have no-slip and no-penetration conditions with zero velocity. The coordinate-system origin is located at the front-face left-bottom corner. The last boundary conditions correspond to the cylinder actuators which enable the control. \rev{The cylinder has a total of two sets of $n_{\rm jets} = 10$ aligned synthetic jets that extend along its entire spanwise dimension, with a spanwise width $L_{\rm jet}/D = 0.4$. This configuration enables having around two actuators per spanwise wavelength, which is known to be of the order of one diameter (mode-B).} Future research could explore the optimal sizes and locations of these actuators for improved performance. Each set is placed at the top and bottom of the cylinder (at $\theta_0^{\rm{top}}=90^{\circ}$ and $\theta_0^{\rm{bottom}}=270^{\circ}$, respectively), defined as independent boundaries. The mass-flow rate can be changed by external actors (the DRL agent in this case, as discussed below). These actuators have an arc length in the $xy$ plane of $\omega = 10^\circ$ and no gap between the jets in $z$ is considered. The jet-velocity profile is defined in terms of the angle $\theta$ and the desired mass-flow rate $Q$ per unit width:

\begin{equation}
\lVert {U_{\rm{jet}}}(Q,\theta)\lVert = Q\frac{\pi}{\rho D \omega }\cos \left ( \frac{\pi}{\omega}(\theta-\theta_0) \right ),
\label{eq:ujet}
\end{equation}

\noindent where $Q = \dot{m}/L_{z}$ and $|\theta-\theta_0|\in[-\omega/2, \omega/2]$, $\dot{m}$ is the mass flow rate. The absolute value of the jet velocity is projected into the $x$ and $y$ axes, since $\lVert {U_{\rm{jet}}} \lVert$ corresponds to the radial cylinder direction. For each pseudo-environment, we set opposite action values within the pair of top and bottom jets \textit{i.e.} $Q_{\rm{90^\circ}}=-Q_{\rm{270^\circ}}$, in order to ensure the global zero net mass flux. An earlier version of this setup was developed in Ref.~\cite{suarez_2023_etmm}.

A conceptually similar control approach was reported in Ref.~\cite{choi_distributed_2005}, in particular the one they denote as \textit{out-phase}. Out-phase approach consists of different constant mass-flow sinusoidal distributions along the spanwise with different wavelengths.

The transition from laminar to the emergence of first three-dimensional instabilities in the cylinder wake occurs within a range of $Re_D=200$ and $300$~\cite{williamson_vortex_1996, bays1993streamwise, bloor1964transition, karniadakis1989frequency, karniadakis1992three, norberg1994experimental}. The motivation of this research is to challenge the control system to discover the optimal strategies as the flow starts to become three-dimensional. By being able to manipulate scales with different spanwise wavelengths in the wake, the DRL-based control can learn effective mechanisms leading to a substantial reduction of the drag in the cylinder.


The numerical simulations are carried out by means of the numerical solver Alya, which is described in detail in Ref.~\cite{vazquez_alya_2016}. The spatial discretization is based on the finite-element method (FEM) and the incompressible Navier--Stokes equations:

\begin{align}
\partial_{t}\bm{u}+(\bm{u}\cdot \nabla)\bm{u}-\nabla \cdot(2\nu \bm{\epsilon})+\nabla p &= \bm{f}, \label{eqn:NS1}\\
\nabla \cdot \bm{u} &= 0,
  \label{eqn:NS2}
\end{align}

\noindent are integrated numerically, where $\bm{\epsilon}$ is a function of the velocity $\bm{u}$ which defines the velocity strain-rate tensor $\bm{\epsilon}=1/2(\nabla \bm{u} + (\nabla\bm{u})^\mathrm{T})$, $\bm{f}$ are the external body forces and $\nu$ is the kinematic viscosity. In Equation (4), the convective term $(\bm{u}\cdot\nabla)\bm{u}$ is expressed as a term conserving energy, momentum and angular momentum~\cite{charnyi_conservation_2017, charnyi_efficient_2019}. For the time discretization, a semi-implicit method is used where the convective term follows a second-order Runge--Kutta scheme and a Crank--Nicholson scheme is used for the diffusive term~\cite{crank_practical_nodate}. To select the appropriate time step, Alya uses an eigenvalue-based time-integration scheme~\cite{article_trias}. Then, for each time step, the numerical solution of these equations is computed. Drag and lift forces ($F_x$ and $F_y$, respectively) are obtained by integration over the cylinder surface $\bm{s}$:

\begin{equation} 
\bm{F}=\int (\bm{\varsigma} \cdot \bm{n}) \cdot \bm{e}_j \rm{d} \bm{s},
  \label{eqn:force}
\end{equation}

\noindent where $\bm{\varsigma}$ is the Cauchy stress tensor, $\bm{n}$ the unit vector normal to the surface and $\bm{e}_j$ is a unit vector with the direction of the main flow velocity for $F_x$ and the perpendicular cross-flow direction to it for $F_y$. 

\subsection*{Multi-agent reinforcement learning (MARL)} \label{sec:MARL}

In the present work, we implemented a deep-reinforcement-learning (DRL) framework using Tensorforce libraries~\cite{schaarschmidt_tensorforce_2017}. DRL is very well suited for unsteady flow-control problems as it provides the possibility to dynamically interact with an environment, being able to dynamically set the actuation based on the varying flow state. We use the proximal-policy-optimization (PPO) algorithm~\cite{schulman2017proximal}, which is a policy-gradient approach based on a surrogate loss function for policy updates to prevent drastic drops in performance. \rev{This algorithm demonstrates robustness, as it is relatively tolerant to initial hyperparameter settings and performs well across a wide variety of RL tasks without requiring extensive tuning. In this context, Ref.~\cite{rabault_accelerating_2019} is referenced, where the effects of parallel environments, action frequency update \(T_a/T_k\), and the smoothing law are studied. Also, Ref.~\cite{rabault_artificial_2019} where in Appendix E, different $s_t$ had consistent results.}

The neural-network architecture consists of two dense layers of 512 neurons each. The batch size, \textit{i.e.} the total number of experiences that the PPO agent uses for each gradient-descent iteration, is set to 80, which is larger than the typical values used in 2D trainings~\cite{varela_deep_2022, rabault_artificial_2019}, but sufficiently small to efficiently update the neural-network parameters. If there is a discrepancy between the batch size and the environments running at the same time there is a risk of wasting information that will not be captured by the agent. \rev{The limitation is that we have 10 actuators per environment, $n_{\rm jets}=10$, and we need 10 streamed experiences which will be synchronized, so we have to work with a total of $n_{\rm jets}\times n_{\rm envs}$ set of experiences.} A streamed experience consists of a set of states, actions, rewards, and the predicted state that the agent expects to achieve. It is denoted as \rev{$(s_t,a_t,r_t,s_t')_{i}$} for each pseudo-environment, and each of the Reynolds numbers under consideration has its own agent and policy. 



\rev{Previous work on 2D cylinders employed single-agent reinforcement learning (SARL) to implement various training stages. However, if the action space needs to handle multiple jets simultaneously---such as in the present 3D cylinder setup with distributed input forcing and a distributed output reward (the so-called DIDO scheme)---SARL is not viable. Unlike SARL, the Multi-Agent Reinforcement Learning (MARL) framework mitigates the curse of dimensionality by exploiting symmetries and training agents in local pseudo-environments. This approach makes high-dimensional control tasks more tractable by breaking them into smaller domains, allowing agents to maximize local rewards effectively. Furthermore, by distributing both the input and output spaces across multiple agents, the exploration space becomes significantly more manageable. This not only accelerates convergence during training but also ensures that the training process remains computationally feasible for such high-dimensional challenges.

It is important to note that earlier attempts using SARL, although not detailed in this study, were unsuccessful in achieving the desired performance—similar to the findings in Rayleigh-Bénard convection reported in Ref. \cite{vignon_effective_2023}. In both cases, SARL struggled with handling global information effectively, leading to suboptimal control strategies. The limitations of the single-agent approach became evident even in simpler scenarios, where it failed to generalize across complex, high-dimensional environments. In contrast, MARL not only overcome these challenges but also demonstrated superior performance by leveraging localized coordination among agents. This ability to process distributed input and optimize distributed rewards makes MARL a much more effective framework, even in cases where SARL had previously been considered feasible.}

Figure~\ref{fig:fig6_sketch}(b) can help to understand the forthcoming explanation of the MARL setup. All the agents share the same neural-network weights, which is a key factor in significantly accelerating the training process. Note that each agent is coupled to a pair of jets that actuate independently from the others through the training process. 

\rev{The observation state \( s_t \) provided to the agent consists of partial pressure values along the domain. This information is composed of three slices, each containing 99 pressure values, which are aligned with the corresponding jet in $z/D$ coordinates and separated by $L_z/30D$. The probes or pressure values are concentrated in the wake and near-cylinder regions, allowing the agent to effectively exploit the spanwise pressure gradients. 

In prior work in Ref.~\cite{suarez_2023_etmm}, various configurations were tested, including changes to the spanwise location and number of slices, to evaluate their impact on the performance of the drag reduction algorithm. These tests were crucial. The configuration chosen for this study was selected because it consistently provided the best overall performance across the evaluated scenarios.}

The total reward $R(t,i_{\rm{jet}})$ defined in Equation (\ref{eq:reward_eq}) is expressed as a sum of the local, $r_{\rm{local}}$, and global, $r_{\rm{global}}$, rewards that correspond to each jet $i_{\rm{jet}}$. The heuristic scalar $K_R$ adjusts the values within the range $[-1,1]$, and $\beta$ balances the local and global rewards; note that a value of $\beta=0.8$ is used in this work. \rev{This means that 80\% of the weight is assigned to the local value, while the remaining 20\% accounts for the global value. Based on our experience, the parameter \( \beta \) also acts as a smoother for the reward signal. If \( \beta \) is too high, the signal can become noisy, whereas a lower value significantly reduces fluctuations. Thus, in terms of learning a policy, the control authority, understood as the ability to influence the behavior of the system given feedback, can also be significantly influenced.

We acknowledge that we did not experimentally test a wide range of \( \beta \) values, as this would have been computationally prohibitive. However, this limitation motivates future research to better understand the sensitivity of \( \beta \) and its role in balancing local and non-local information. Importantly, \( \beta \) plays a key role in facilitating coordination between agents, enabling neighboring agents to exchange meaningful information. This coordination is critical because what is beneficial for local performance may sometimes conflict with global objectives, and vice versa. By defining \( \beta=0.8 \), the framework seeks to find a balance that harmonizes local and global priorities.} 

The rewards $r_t$, defined in Equation (\ref{eq:reward_cd}), are functions of the aerodynamic force coefficients $C_D$ and $C_L$ \rev{(note that $\overline{C_{D_{\rm{b}}}}$} is the uncontrolled averaged drag in a stationary state). The user-defined parameter $\alpha$ is a lift penalty, and in this study, we considered $\alpha=0.6$, which provides a good trade-off between ensuring symmetric strategies without excessively restricting the exploration process. The latter is essential to avoid undesired asymmetric strategies that favor a reduction of the component parallel to the incident velocity (drag) towards the perpendicular one (positive or negative lift). This phenomenon is commonly referred to as the axis-switching phenomenon.

\begin{align}
    \label{eq:reward_eq} R(t,i_{\rm{jet}})=K_R \left [\beta r_{\rm{local}}(t,i_{\rm{jet}}) + (1-\beta)r_{\rm{global}}(t) \right ], \\
    \label{eq:reward_cd} r(t,i_{\rm{jet}})=\overline{C_{D_{\rm{b}}}}-C_D(t,i_{\rm{jet}})-\alpha\vert C_L(t,i_{\rm{jet}})\vert,\\
    \text{where} \quad C_D=\frac{2 F_x}{\rho A_f U_{\infty}^{2}}  \quad \text{and} \quad C_L=\frac{2 F_y}{\rho A_f U_{\infty}^{2}}.
\end{align} 

\noindent \rev{The aerodynamic forces involve the frontal area $A_f=DL_z$ from the local pseudo-environment surfaces for $C_{D_{\rm{local}}}$ and the whole cylinder for $C_{D_{\rm{global}}}$.}

The interactions between the agent and the physical environment are denoted as actions $a_t$, and they influence the system during $T_a$ time units. We update the jet boundary conditions using Equation (\ref{eq:ujet}). The shift in time between actions, $Q_{t}\rightarrow Q_{t+1}$ is managed through an exponential function. The smooth transition diminishes the appearance of sudden discontinuities which can spoil a training process. 

\begin{table}[H]
        \centering
        \resizebox{0.60\textwidth}{!}{
        \begin{tabular}{@{}ll@{}}
            \toprule
            \textbf{Parameter} & \textbf{Value} \\ \midrule \midrule
            $n_{\rm jets}$  & 10 \\
            $s_t$ size  & 297 (99 in 3 $xy$-slices) \\
            $s_t$ variable & Pressure \\
            $Q_{\rm max}$ & 0.176 \\
            Reward scalar $K_R$ & 5 \\
            Lift penalty $\alpha$ & 0.6 \\
            Reward local weight $\beta$ & 0.8 \\
            Action duration $T_a [tU_\infty/D]$ & 0.25\\
            Actions per episode & 120 \\
            Time-smoothing function  & Exponential \\
            Batch size $M$ & 60 \\
            Epoch for optimizer & 25 \\
            Strategy optimize policy & PPO \\
            Architecture (networks) & 2 \\
            Network 1 \& 2 (type) & Fully connected layer \\
            Network 1 \& 2 (size) & 512 neurons \\
            Learning rate $\alpha_\theta$ and $\alpha_\phi$ & 0.001 \\
            Likelihood ratio clipping $\epsilon$ & 0.2 \\
            Discount factor $\gamma$ & 0.99 \\
            Optimizer type & Adam \\
            Entropy regularization $\lambda$ & 0.01 \\                
            \bottomrule
        \end{tabular}}
        \caption{\textbf{\rev{Summary of the main parameters for the present DRL framework.}}}
        \label{tab:table_MARL}
\end{table}

\rev{The DRL library outputs values in the range \( a_t \in [-1, 1] \), requiring rescaling as \( Q = a_t Q_{\rm{max}} \) to introduce the magnitude of actuation later in Equation~\ref{eq:ujet}. To balance exploration, learning efficiency, and numerical stability, $Q_{\rm{max}} = 0.176$ was selected based on our experience with DRL for flow control. This value ensures meaningful exploration without introducing excessively large actuations that could destabilize the learning process or the CFD solver. If $Q_{\rm{max}}$ were set too small, the exploration space would be overly constrained, potentially leading to premature policy saturation and suboptimal solutions. On the other hand, excessively large $Q_{\rm{max}}$ values could result in an expanded exploration space that delays convergence or imposes boundary conditions that challenge the CFD solver’s stability. Notably, $Q_{\rm{max}} = 0.176$ corresponds to twice the values used in the 2D cylinder setups~\cite{varela_deep_2022}, reflecting adjustments for the current configuration and objectives.}

\rev{Certain parameters in the DRL configuration are closely tied to the fluid-mechanics problem under consideration. The episode duration is specifically defined to include at least six vortex-shedding periods ($T_k=1/St$).} We set $T_a<0.05T_k$, based on the experience gathered with previous studies~\cite{tang_robust_2020, rabault_artificial_2019}. This allows sufficient time between actions to produce an effect on the flow. Note that if the time between actions is too short, there will be noise in the training process and it will become difficult to converge. On the other hand, if this is too large the agent will not be able to control the smaller-scale structures associated with shorter time scales. Thus, a total of 120 actuations per episode is deemed sufficient for evaluating the cumulative reward. It is noteworthy that each episode starts from an uncontrolled converged state of the problem. This corresponds to what happens during training, but when we evaluate the DRL model in exploitation mode (also denoted as a deterministic mode), we make the episodes 4 times longer to ensure statistical convergence.

\rev{The Table~\ref{tab:table_MARL} collects and summarizes all the main parameters required to set up this DRL framework, which is coupled with a CFD solver, many of which were discussed in this section. Note that the agent hyperparameters like $\epsilon$ or $\alpha_\theta$, are also included but are not discussed in detail in this study.}

Note that we also compare the DRL-based control with results from the classical periodic control. The latter is chosen with the same jet flow rate as that of the DRL, and the frequency is chosen based on a parametric analysis of the frequency around the vortex-shedding frequency of the wake. We selected the frequency yielding the highest drag reduction.






Extensive work documented in Ref.~\cite{suarez_2023_etmm} was carried out to adjust the MARL framework and the communications setup. For instance, the definition of $s_t$ is the result of a compromise between computational practicality and physical relevance. The spanwise wavelength of the structures in the wake also helped to define the spacing of the $xy$ planes defining the system state. Note that the number of data points used for this state correlates with the number of weights calculated for the first fully connected layer of the neural network.    

\section*{Acknowledgments}
This study was enabled by resources provided by the National Academic Infrastructure for Supercomputing in Sweden (NAISS) at PDC, KTH Royal Institute of Technology. R.V. acknowledges financial support from ERC grant no.2021-CoG-101043998, DEEPCONTROL. Views and opinions expressed are however those of the author(s) only and do not necessarily reflect those of the European Union or the European Research Council. Neither the European Union nor the granting authority can be held responsible for them.

\label{sec:acknowledges}

\section*{Author Contributions}

\textbf{Suárez, P.:}  Methodology, software, validation, investigation, writing - original draft and visualization. \textbf{Álcantara-Ávila, F., Rabault, J., Miró, A. \& Font, B. :} Methodology, software, and writing - review \& editing. \textbf{Lehmkuhl, O. :} Funding acquisition, supervision, and writing - review \& editing. \textbf{Vinuesa, R.:} Conceptualization, project definition, methodology, resources, writing - original draft, supervision, project administration and funding acquisition.

\section*{Data availability}
The data and codes used to produce this study will be made available for open access as soon as the article is published.

\newpage

\section*{Appendix 1: Case validation and convergence}
\rev{A grid-independence study was conducted to select an optimal mesh that balances low computational cost and accurate physical representation. This step is particularly important when working with the DRL framework due to its high computational  demands and the necessity of parallelizing multiple simulations simultaneously. 
The study involves iterating over a parametrized mesh, adjusting specific values to control the resolution in the \(xy\)-plane and the number of uniformly spaced  elements along the \(z\)-axis. Note that lengths $L_x/D=30$ and $L_y/D=15$ are the same for all cases. The schematic illustrates the placement of these elements and the parameters that can be adjusted. The table lists the details of each mesh—ranging from coarse to finer resolutions—and the resulting statistics for the quantities of interest, such as \(\overline{C_D}\), \(\overline{C_{D_{\rm RMS}}}\), and \(\overline{C_{L_{\rm RMS}}}\), calculated over the entire surface. To ensure temporal convergence, 200 time units were found to be more than sufficient to achieve it for this $Re_D$ range.

The results obtained for the mean drag coefficient (\(\overline{C_D}\)) of a circular cylinder in the Reynolds-number range between  100 to 400—\(\overline{C_D} = 1.4\) at \(Re_D = 100\), \(1.38\) at \(Re_D = 200\), \(1.3\) at \(Re_D = 300\), and \(1.21\) at \(Re_D = 400\)—are in good agreement with the ranges reported in the literature (\(\overline{C_D} \approx 1.2 - 1.4\) at \(Re_D = 100\), \(1.1 - 1.3\) at \(Re_D = 200\), \(1.0 - 1.2\) at \(Re_D = 300\), and \(1.0 - 1.2\) at \(Re_D = 400\)). Some  scatter in the data is observed, which can be attributed to the coexistence of two modes in this transition regime, as discussed in the works of \cite{tritton_1959, schlichting2016boundary, white2006viscous}. Regarding the Strouhal number $St$, the literature exhibits  less scattering across \(Re_D\) and better alignment with the reported values in the  experimental and numerical literature. The first columns in the table in Figure~\ref{fig:fig2}(c) illustrate this agreement, comparing \(St\) values from the literature (Williamson, 1996 \cite{williamson_vortex_1996}) with the present statistics. Furthermore, the \(St\) values enable us to associate modes with our results as well.
}

\begin{figure}[H]
    \centering
    \includegraphics[width=0.8\textwidth]{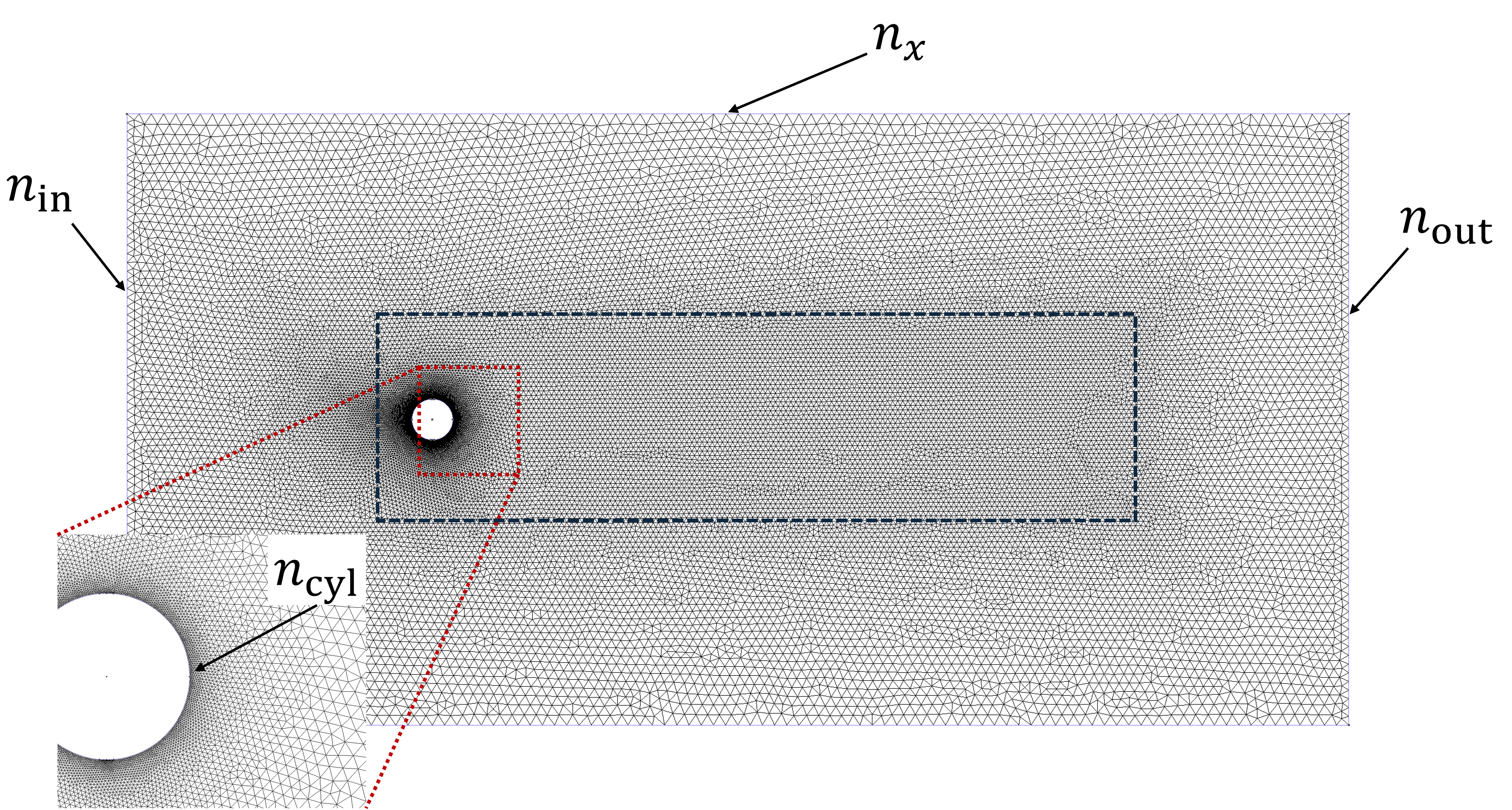}\\
    \caption{\rev{Schematic representation of the parametric mesh used in this study. Note the main number of elements for inlet ($n_{\rm in}$), cylinder ($n_{\rm cyl}$), outlet ($n_{\rm out}$), top and bottom faces ($n_{x}$)). The dashed square indicates the area designed to improve the homogeneity of element sizes around the wake. There is a zoom over the cylinder to note the refinement done close to surface. The element distribution is uniform in spanwise and $n_z$ is not displayed.}}
\label{fig:fig5}
\end{figure}

\begin{table}[H]
\centering
\resizebox{1\textwidth}{!}{
\begin{tabular}{cccccccccccccc}
Mesh   & $L_z/D$   & \multicolumn{1}{l}{$n_{\rm cyl}$} & \multicolumn{1}{l}{$n_{\rm x}$} & \multicolumn{1}{l}{$n_{\rm in}$} & \multicolumn{1}{l}{$n_{\rm out}$} & \multicolumn{1}{l}{$n_{\rm z}$} & \multicolumn{1}{l}{$\Delta z$} & \multicolumn{1}{l}{$min(\Delta r)$} & \multicolumn{1}{l}{Elements} & \multicolumn{1}{l}{Elements/xy} & $\overline{C_D}$ & $\overline{C_{D_{\rm RMS}}}$ & $\overline{C_{L_{\rm RMS}}}$  \\
\hline
\hline
C                & 4                  & 200                         & 60                       & 50                           & 50                            & 110                      & 0,0364                       & 0,0096                                                  & 4,31E+06                     & 3,92E+05                        & 1,215  & 0,038   & 0,336   \\
M1      & 4                             & 240                         & 70                       & 50                           & 50                            & 130                      & 0,0308                       & 0,0050                                                    & 5,29E+06                     & 4,07E+05                        & 1,233  & 0,040   & 0,347   \\
M2        & 6                     & 240                         & 70                       & 50                           & 50                            & 195                      & 0,0308                       & 0,0050                                                   & 7,94E+06                     & 4,07E+05                        & 1,234  & 0,038   & 0,344   \\
M3       & 12                       & 240                         & 70                       & 50                           & 50                            & 360                      & 0,0333                       & 0,0050                                                    & 1,45E+07                     & 4,03E+05                        & 1,233  & 0,035   & 0,342   \\
F1          & 4                        & 320                         & 100                      & 70                           & 70                            & 130                      & 0,0308                       & 0,0042                                                   & 6,62E+06                     & 5,09E+05                        & 1,231  & 0,047   & 0,327   \\
F2            & 4                     & 380                         & 100                      & 70                           & 70                            & 150                      & 0,0267                       & 0,0025                                                    & 7,94E+06                     & 5,29E+05                        & 1,236  & 0,041   & 0,380   \\
F3             & 4                     & 380                         & 110                      & 80                           & 80                            & 150                      & 0,0267                       & 0,0025                                                   & 9,03E+06                     & 6,02E+05                        & 1,218  & 0,041   & 0,338  
\end{tabular}
}
\caption{ \rev{Summary of grid independence study for $Re_D=400$ of seven different meshes: coarse (C), medium (M) with different spanwise lengths, and fine (F1, F2, F3). All quantities are averaged over the last 200 time units.}}
\end{table}

\newpage

\section*{Appendix 2: Periodic control study}

\rev{This appendix aims to briefly introduce the study conducted to serve as a reference for a classical active control approach and be compared with the DRL performance. For this purpose, we opted for a periodic control (PC), sweeping through frequencies \(f_c\lvert_{\rm PC}\) around the previously known \(St \lvert_{\rm Uncontrolled}\) values for each case, and increasing \(Q\) up to \(Q_{\rm max}=0.176\) as a reference for maximum amplitude, since this is allowed by the DRL framework for exploration. Thus, the actuation simplifies to a flow with a profile given by Equation(\ref{eq:ujet}) in the \(xy\)-plane, extruded uniformly in the \(z\)-direction. In this case, however, \(Q(f_c\lvert_{\rm PC},t) = Q_{\rm max} \sin(2\pi f_c\lvert_{\rm PC} t)\).

Figure 8 shows the heatmaps of drag reduction for each of the \(Re_D\) values. It can be observed that  it is difficult to identify a clear pattern, especially since we only have two degrees of freedom in this case. It appears that the system is highly sensitive to even small changes in \(f_c\) or \(Q_{\rm max}\). The cases with the best drag reduction were selected for comparison in the study and serve as the reference for classic or periodic control.

Building on these results and the success of the DRL approach, we investigated whether the trained models could also be simplified using a spanwise extruded profile. In this approach, we can simply select the dominant \(f_c \lvert_{\rm DRL}\) from the DRL actions spectra—refer to the table in Figure~\ref{fig:fig2}(c)—and perform a sweep around \(Q_{\rm max}\) again, but now with values close to those obtained  when running the deterministic cases—refer to Figure~\ref{fig:fig3_mfr}(a).

The results are enlightening: it is not possible to reduce the solution we already know from DRL to a simpler control. Achieving comparable drag reductions requires closed-loop control with the effective ability to transition to a stabilized control state. Furthermore, we can emphasize that beyond the transition, the exploitation of the three-dimensionalities becomes more necessary, and the performance of these simpler systems becomes ineffective.}

\begin{figure}[]
    \centering
    \begin{minipage}{\textwidth}  
    
        \centering
        \resizebox{0.65\textwidth}{!}{
        \begin{tabular}{cc||ccccc}
        
        \multicolumn{2}{c}{$f_c\lvert_{\rm PC}/St \lvert_{\rm Uncontrolled}$} & 0,7                            & 0,85                           & 1                              & 1,15                           & 1,3                            \\
        \multicolumn{2}{c}{$f_c\lvert_{\rm PC}$}            & 0,1148                         & 0,1394                         & 0,164                          & 0,1886                         & 0,2132                         \\
        \hline \hline
                                       & 0,0088       & \cellcolor[HTML]{D9EAD3}-0,051 & \cellcolor[HTML]{D9EAD3}-0,463 & \cellcolor[HTML]{FCE5CD}0,557  & \cellcolor[HTML]{D58886}19,010 & \cellcolor[HTML]{D9EAD3}-0,174 \\
                                       & 0,0117       & \cellcolor[HTML]{D9EAD3}-0,217 & \cellcolor[HTML]{D9EAD3}-1,549 & \cellcolor[HTML]{FCE5CD}1,708  & \cellcolor[HTML]{D9EAD3}-0,051 & \cellcolor[HTML]{D9EAD3}-0,275 \\
                                       & 0,0176       & \cellcolor[HTML]{D9EAD3}-0,825 & \cellcolor[HTML]{D9EAD3}-1,042 & \cellcolor[HTML]{FFCCC9}4,242  & \cellcolor[HTML]{D9EAD3}-0,101 & \cellcolor[HTML]{D9EAD3}-0,029 \\
        \multirow{-4}{*}{$Q_{\rm max}$}       & 0,0528       & \cellcolor[HTML]{B6D7A8}\textbf{-4,850} & \cellcolor[HTML]{FFCCC9}4,365  & \cellcolor[HTML]{D58886}15,825 & \cellcolor[HTML]{D58886}15,825 & \cellcolor[HTML]{D9EAD3}-2,512
        \end{tabular}
        }
        \subcaption*{(a) $Re_D=100$}

        \centering
        \resizebox{0.65\textwidth}{!}{
        \begin{tabular}{cc||ccccc}
        
        \multicolumn{2}{c}{$f_c\lvert_{\rm PC}/St \lvert_{\rm Uncontrolled}$} & 0,7                            & 0,85                           & 1                              & 1,15                           & 1,3                            \\
        \multicolumn{2}{c}{$f_c\lvert_{\rm PC}$}            & 0,1302                        & 0,1581                         & 0,186                          & 0,214                         & 0,2418 \\
        \hline \hline
                               & 0,0088       & \cellcolor[HTML]{D9EAD3}-0,900          & \cellcolor[HTML]{D9EAD3}-0,755 & \cellcolor[HTML]{D9EAD3}-1,919 & \cellcolor[HTML]{FCE5CD}0,191  & \cellcolor[HTML]{D9EAD3}-0,318 \\
                               & 0,0117       & \cellcolor[HTML]{D9EAD3}-0,306          & \cellcolor[HTML]{D9EAD3}-1,846 & \cellcolor[HTML]{D9EAD3}-0,100 & \cellcolor[HTML]{FCE5CD}1,574  & \cellcolor[HTML]{FCE5CD}0,701  \\
                               & 0,0176       & \cellcolor[HTML]{D9EAD3}-1,409          & \cellcolor[HTML]{B6D7A8}-3,956 & \cellcolor[HTML]{FCE5CD}1,785  & \cellcolor[HTML]{FFCCC9}2,163  & \cellcolor[HTML]{FCE5CD}0,410  \\
\multirow{-4}{*}{$Q_{\rm max}$}       & 0,0528       & \cellcolor[HTML]{B6D7A8}\textbf{-8,322} & \cellcolor[HTML]{FFCCC9}2,592  & \cellcolor[HTML]{D58886}17,654 & \cellcolor[HTML]{D58886}14,234 & \cellcolor[HTML]{FCE5CD}0,410 

        \end{tabular}
        }
        \subcaption*{(b) $Re_D=200$}

        \centering
        \resizebox{0.65\textwidth}{!}{
        \begin{tabular}{cc||ccccc}
        
        \multicolumn{2}{c}{$f_c\lvert_{\rm PC}/St \lvert_{\rm Uncontrolled}$} & 0,7                            & 0,85                           & 1                              & 1,15                           & 1,3                            \\
        \multicolumn{2}{c}{$f_c\lvert_{\rm PC}$}            & 0,144                         & 0,175                         & 0,206                          & 0,237                         & 0,268                         \\
        \hline \hline
                               & 0,0088       & \cellcolor[HTML]{D9EAD3}-2,421                         & \cellcolor[HTML]{D9EAD3}-0,908 & \cellcolor[HTML]{FCE5CD}5,900  & \cellcolor[HTML]{D9EAD3}-1,664                         & \cellcolor[HTML]{D9EAD3}-1,740 \\
                               & 0,0117       & \cellcolor[HTML]{D9EAD3}-2,421                         & \cellcolor[HTML]{B6D7A8}-3,933 & \cellcolor[HTML]{FCE5CD}8,926  & \cellcolor[HTML]{D9EAD3}-0,908                         & \cellcolor[HTML]{FCE5CD}0,756  \\
                               & 0,0176       & \cellcolor[HTML]{B6D7A8}-4,690 & \cellcolor[HTML]{B6D7A8}\textbf{-6,959} & \cellcolor[HTML]{F4CCCC}14,221 & \cellcolor[HTML]{FCE5CD}5,144  & \cellcolor[HTML]{D9EAD3}-1,664 \\
\multirow{-4}{*}{$Q_{\rm max}$}        & 0,0528       & \cellcolor[HTML]{D9EAD3}-0,908                         & \cellcolor[HTML]{F4CCCC}11,952 & \cellcolor[HTML]{D58886}21,785 & \cellcolor[HTML]{D9EAD3}-0,151                         & \cellcolor[HTML]{B6D7A8}-5,068
        \end{tabular}
        }
        \subcaption*{(c) $Re_D=300$}

                \centering
        \resizebox{0.65\textwidth}{!}{
        \begin{tabular}{cc||ccccc}
        
        \multicolumn{2}{c}{$f_c\lvert_{\rm PC}/St \lvert_{\rm Uncontrolled}$} & 0,7                            & 0,85                           & 1                              & 1,15                           & 1,3                            \\
        \multicolumn{2}{c}{$f_c\lvert_{\rm PC}$}            & 0,141                        & 0,172                         & 0,202                          & 0,232                         & 0,263132                         \\
        \hline \hline
                                & 0,0088       & \cellcolor[HTML]{D9EAD3}-2,526                         & \cellcolor[HTML]{D9EAD3}-1,973 & \cellcolor[HTML]{FCE5CD}6,314  & \cellcolor[HTML]{D9EAD3}-6,077                         & \cellcolor[HTML]{D9EAD3}-1,579                         \\
                               & 0,0117       & \cellcolor[HTML]{D9EAD3}-4,499                         & \cellcolor[HTML]{B6D7A8}\textbf{-9,629} & \cellcolor[HTML]{FCE5CD}8,019  & \cellcolor[HTML]{D9EAD3}-5,525                         & \cellcolor[HTML]{D9EAD3}-3,552                         \\
                               & 0,0176       & \cellcolor[HTML]{B6D7A8}-7,751 & \cellcolor[HTML]{B6D7A8}-7,893 & \cellcolor[HTML]{F4CCCC}10,497 & \cellcolor[HTML]{D9EAD3}-5,975                         & \cellcolor[HTML]{D9EAD3}-5,051                         \\
\multirow{-4}{*}{$Q_{\rm max}$}       & 0,0528       & \cellcolor[HTML]{D9EAD3}-0,710                         & \cellcolor[HTML]{D58886}18,564 & \cellcolor[HTML]{D58886}37,332 & \cellcolor[HTML]{FCE5CD}10,576 & \cellcolor[HTML]{D9EAD3}-5,462      
        \end{tabular}
        }
        \subcaption*{(d) $Re_D=400$}
        
    \end{minipage}
    
    \caption{\rev{Drag reduction percentage heat map (green is beneficial and red are degrading cases), $\Delta \overline{C_D}[\%]=100\times(\overline{C_D \lvert_{\rm controlled}}/\overline{C_D \lvert_{\rm uncontrolled}}-1)$, summarizing the results of periodic control across a sweep of different frequencies, $f_c\lvert_{\rm PC}$, and amplitudes, $Q_{\rm max}$, for various $Re_D$. Values in bold represent the best cases selected for comparison against the DRL results.}}
    \label{fig:app_2_fc}
\end{figure}

\begin{figure}[]
    
    \begin{minipage}{\textwidth}  
    \centering
    \resizebox{0.9\textwidth}{!}{
        \begin{tabular}{c||cccccccc}
        $Q_{\rm max}$                             & 0,0044                         & 0,0059                         & 0,0088                         & 0,0117                         & 0,0176                         & 0,0528                         & 0,0880                         & 0,1232                         \\
        $\Delta \overline{C_D} [\%]$ & \cellcolor[HTML]{D9EAD3}-0,101 & \cellcolor[HTML]{D9EAD3}-0,825 & \cellcolor[HTML]{D9EAD3}-1,549 & \cellcolor[HTML]{B6D7A8}-4,734 & \cellcolor[HTML]{D9EAD3}-1,339 & \cellcolor[HTML]{FFCCC9}11,235 & \cellcolor[HTML]{FFCCC9}19,314 & \cellcolor[HTML]{FFCCC9}20,530
        \end{tabular}}
    \subcaption*{(a) $Re_D=100$ with control frequency set from known DRL, $f_c\lvert_{\rm DRL}=0.154$.}

    \resizebox{0.9\textwidth}{!}{
        \begin{tabular}{c||cccccccc}
$Q_{\rm max}$                              & 0,0044                         & 0,0059                         & 0,0088                         & 0,0117                         & 0,0176                         & 0,0528                        & 0,0880                         & 0,1232                                    \\
 $\Delta \overline{C_D} [\%]$  & \cellcolor[HTML]{D9EAD3}-6,139 & \cellcolor[HTML]{D9EAD3}-8,191 & \cellcolor[HTML]{B6D7A8}-9,777 & \cellcolor[HTML]{D9EAD3}-6,139 & \cellcolor[HTML]{D9EAD3}-4,931 & \cellcolor[HTML]{FFCCC9}9,869 & \cellcolor[HTML]{FFCCC9}26,603 & \cellcolor[HTML]{FFCCC9}52,797 
\end{tabular}}
    \subcaption*{(b) $Re_D=200$ with control frequency set from known DRL, $f_c\lvert_{\rm DRL}=0.173$.}

    \resizebox{0.9\textwidth}{!}{
    \begin{tabular}{c||ccccccccc}
    $Q_{\rm max}$                               & 0,0009                         & 0,0018                         & 0,0035                         & 0,0044                         & 0,0059                         & 0,0088                        & 0,0117                        & 0,0176          & 0,0528        \\
    $\Delta \overline{C_D} [\%]$ & \cellcolor[HTML]{D9EAD3}-3,555 & \cellcolor[HTML]{D9EAD3}0,076 & \cellcolor[HTML]{D9EAD3}-1,740 & \cellcolor[HTML]{D9EAD3}-0,076 & \cellcolor[HTML]{D9EAD3}0,681 & \cellcolor[HTML]{D9EAD3}-1,362 & \cellcolor[HTML]{D9EAD3}-1,740 & \cellcolor[HTML]{B6D7A8}-5,673  & \cellcolor[HTML]{FFCCC9}4,0 
    \end{tabular}}
    \subcaption*{(c) $Re_D=300$ with control frequency set from known DRL, $f_c\lvert_{\rm DRL}=0.158$.}

    \resizebox{0.9\textwidth}{!}{
        \begin{tabular}{c||ccccccccc}
$Q_{\rm max}$                              & 0,0009                         & 0,0018                         & 0,0035                         & 0,0044                         & 0,0059                        & 0,0088                        & 0,0117                        & 0,0176                         &0.0528\\
$\Delta \overline{C_D} [\%]$ & \cellcolor[HTML]{D9EAD3}-0,868& \cellcolor[HTML]{D9EAD3}-0,789& \cellcolor[HTML]{D9EAD3}-2,605& \cellcolor[HTML]{D9EAD3}-3,552& \cellcolor[HTML]{D9EAD3}-3,315& \cellcolor[HTML]{D9EAD3}-1,973& \cellcolor[HTML]{D9EAD3}-5,028& \cellcolor[HTML]{B6D7A8}-7,545 &\cellcolor[HTML]{FFCCC9}16,5\end{tabular}}
    \subcaption*{(d) $Re_D=400$ with control frequency set from known DRL, $f_c\lvert_{\rm DRL}=0.171$.}

\end{minipage}
    \caption{\rev{Drag reduction percentage, $\Delta \overline{C_D}[\%]$, summarizing the results of periodic control across a sweep of different amplitudes, $Q_{\rm max}$. Note that this control is with fixed frequency taking the ones obtained from DRL, $f_c\lvert_{\rm DRL}$, for each studied $Re_D$. Tables are color-coded to indicate performance: beneficial cases are shown in green, while degrading cases are shown in red.}}
    \label{fig:app_2_fc_drl_reference}
\end{figure}

\bibliographystyle{elsarticle-num-names}
\bibliography{lib_pol}

\end{document}